\documentclass[10pt,a4paper]{article}

\usepackage{latexsym,amsfonts,amsmath,amssymb,mathrsfs,url,color,amsthm}
\usepackage{graphicx,cite}

\usepackage[dvipdfm,
linkbordercolor={1 0 0},
citebordercolor={0 1 0},
urlbordercolor={0 1 1}]{hyperref}

\newcommand{\rd}{\,\mathrm{d}}
\newcommand{\bsA}{\boldsymbol{A}}
\newcommand{\EE}{\mathbb{E}}
\newcommand{\NN}{\mathbb{N}}
\newcommand{\RR}{\mathbb{R}}

\newcommand{\EVPI}{\mathrm{EVPI}}
\newcommand{\EVPPI}{\mathrm{EVPPI}}

\newcommand{\all}{\mathrm{all}}
\newcommand{\coupled}{\mathrm{coupled}}
\newcommand{\single}{\mathrm{single}}

\begin{document}

\title{Unbiased Monte Carlo estimation for the expected value of partial perfect information\thanks{The work of T.~G. is supported by JSPS Grant-in-Aid for Young Scientists No.~15K20964 and Arai Science and Technology Foundation.}}

\author{Takashi Goda\thanks{Graduate School of Engineering, The University of Tokyo, 7-3-1 Hongo, Bunkyo-ku, Tokyo 113-8656, Japan. ({\tt goda@frcer.t.u-tokyo.ac.jp})}}

\date{\today}

\maketitle

\begin{abstract}
The expected value of partial perfect information (EVPPI) denotes the value of eliminating uncertainty on a subset of unknown parameters involved in a decision model. The EVPPI can be regarded as a decision-theoretic sensitivity index, and has been widely used for identifying relatively important unknown parameters. It follows from Jensen's inequality, however, that the standard nested Monte Carlo computation of the EVPPI results in biased estimates. In this paper we introduce two unbiased Monte Carlo estimators for the EVPPI based on multilevel Monte Carlo method, introduced by Heinrich (1998) and Giles (2008), and its extension by Rhee and Glynn (2012, 2015). Our unbiased estimators are simple and straightforward to implement, and thus are of highly practical use. Numerical experiments show that even the convergence behaviors of our unbiased estimators are superior to that of the standard nested Monte Carlo estimator.
\end{abstract}
\emph{Keywords}: value of information, expected value of partial perfect information, unbiased estimation, multilevel Monte Carlo

\section{Introduction}\label{sec:1}
Since introduced by Howard \cite{Howard66}, the concept of the expected value of information has long been studied in the context of decision analysis \cite{Delquie08,Raiffa68,SWR89} and applied to various areas, such as medical decision making \cite{BKOC07,MAPMJRW14}, environmental science \cite{BSLL14,DTSB96} and petroleum engineering \cite{BGMPW08,BBL09,NGTS16,Sato11}.
The expected value of information is defined as the expected increase in monetary value brought from reducing some degree of uncertainty on unknown parameters involved in a decision model by obtaining additional information.
There are several definitions of the expected value of information depending on the type of information, which includes perfect information, partial perfect information and sample information.
In particular, the expected value of partial perfect information (EVPPI), or sometimes called the partial expected value of perfect information, denotes the value of eliminating uncertainty on a subset of unknown parameters completely, and has been advocated and used as a decision-theoretic sensitivity index for identifying relatively important unknown parameters \cite{Claxton99,FH98,Oakley09}.

For many problems encountered in practice, calculating the EVPPI analytically is not possible.
The simplest and most often-used method to approximately evaluate the EVPPI is the nested Monte Carlo computation \cite{FH98,BKOC07,OBTC10}.
As pointed out in \cite{BKOC07,OBTC10}, however, the standard nested Monte Carlo computation of the EVPPI results in biased estimates, which directly follows from Jensen's inequality.
Moreover, it can be inferred from \cite[Section~2]{AG16} that the standard nested Monte Carlo computation cannot achieve the square-root convergence rate in the total computational budget.
In fact, the author of this paper empirically observed a deteriorated convergence rate for a simple toy problem in \cite{NGTS16}.
Therefore, an unbiased and efficient computation of the EVPPI might be of particular interest to practitioners.
In this line of investigation, there have been some recent attempts to construct such computational algorithms \cite{BKOC07,CO08,MAPMJRW14,Oakley09,SBZNM13,SO13}.
As far as the author knows, however, every algorithm proposed in the literature has its own restrictions, for instance, on a decision model, and there is no general algorithm with mild assumptions.

In this paper we construct general unbiased Monte Carlo estimators for the EVPPI as well as the expected value of perfect information (EVPI).
Our estimators for the EVPPI on a certain subset of unknown parameters only assume that i.i.d.\ random sampling from the conditional distribution of the complement of unknown parameters should be possible.
If this is not the case, it might be necessary to incorporate Markov chain Monte Carlo sampling into our estimators, although such an investigation is beyond the scope of this paper.
For a decision model which satisfies the above assumption, our estimators are quite simple and straightforward to implement.

Our approach to construct unbiased estimators is based on the multilevel Monte Carlo (MLMC) method, which was first introduced by Heinrich \cite{Heinrich98} for parametric integration and by Giles \cite{Giles08} for path simulation, and was later extended by Rhee and Glynn \cite{RG12,RG15}.
We refer to \cite{Giles15} for a state-of-the-art review on the MLMC method.
The idea of the MLMC method can be simply described as follows:
For a dimension $s$, let $f\in L^2([0,1]^s)$, and $f_1,f_2,\ldots\in L^2([0,1]^s)$ be a sequence of functions which approximates $f$ with increasing accuracy (in the $L^2$ norm) but also with increasing computational cost.
We denote by $I(f)$ the true integral of $f$, i.e.,
  \begin{align*}
   I(f) := \int_{[0,1]^s}f(x)\rd x.
  \end{align*}
The naive Monte Carlo computation chooses $N$ points $x_1,\ldots,x_N$ independently and randomly from $[0,1]^s$ to approximate $I(f)$ by the average
  \begin{align*}
   I_{\text{MC}}(f) = \frac{1}{N}\sum_{n=1}^{N}f(x_n) \quad \text{or}\quad I_{\text{MC}}(f_L) = \frac{1}{N}\sum_{n=1}^{N}f_L(x_n) ,
  \end{align*}
for some $L$.
Note that the former is an unbiased estimator of $I(f)$ whereas the latter contains the bias $I(f)-I(f_L)$.
The MLMC method, on the other hand, uses the telescopic representation $f=f_1+(f_2-f_1)+(f_3-f_2)+\cdots$, and then each term is independently approximated by the naive Monte Carlo computation, i.e.,
  \begin{align*}
   I_{\text{MLMC}}(f) = \sum_{l=1}^{\infty}\frac{1}{N_l}\sum_{n=1}^{N_l}(f_{l}-f_{l-1})(x_{l,n}) ,
  \end{align*}
where we set $f_{0}\equiv 0$ and $N_1+N_2+\cdots = N$.
For the level $l$ such that $N_l=0$, the corresponding average is set to 0.
The original MLMC method in \cite{Giles08} considers the case $N_{L+1}=N_{L+2}=\cdots =0$, that is, the telescopic representation of $f$ is truncated up to $L$ terms.
The resulting estimator contains the bias $I(f)-I(f_L)$.
The extended MLMC method in \cite{RG12,RG15} introduces a probability mass function $p$ such that $p(l)>0$ for all $l\in \NN$, where $\NN$ denotes the set of positive integers, and considers the \emph{single term estimator}
  \begin{align*}
   I_{\text{MLMC}}(f) = \frac{1}{N}\sum_{n=1}^{N}\frac{(f_{l_n}-f_{l_n-1})(x_n)}{p(l_n)} ,
  \end{align*}
or the \emph{coupled sum estimator}
  \begin{align*}
   I_{\text{MLMC}}(f) = \frac{1}{N}\sum_{n=1}^{N}\sum_{l=1}^{l_n}\frac{(f_{l}-f_{l-1})(x_n)}{\sum_{k=l}^{\infty}p(k)} ,
  \end{align*}
where $l_1,\ldots,l_N$ and $x_1,\ldots,x_N$ are chosen independently and randomly according to $p$ and $U([0,1]^s)$, respectively.
These estimators are shown to be unbiased \cite{Giles15,RG12}.
In this setting, the superiority of the MLMC method over the naive Monte Carlo method depends on the balance between the growth rate of the computational costs for $f_1,f_2,\ldots$ and the decay rate of the variances of $f_1-f_0, f_2-f_1, \cdots$.

An application of the MLMC method to the nested Monte Carlo computation in a different context has been done, for instance, in \cite{BHR15} and also mentioned in \cite[Section~9]{Giles15}.
However, the MLMC method has never been applied to computations of the expected value of information.
In this paper, we show that the framework of the MLMC method actually fits quite well into constructing unbiased estimators both for the EVPI and the EVPPI.
Because of their simplicity and efficiency, we believe that our unbiased estimators will be one of the most standard choices particularly for evaluating the EVPPI.
Finally, it should be remarked that an unbiased estimator for optimization of expectations has been constructed very recently by Blanchet and Glynn \cite{BG15} in a general context, whose main approach is commonly used in this paper. 

The remainder of this paper is organized as follows.
In the next section, we introduce the definitions of the EVPI and the EVPPI, and then discuss the standard nested Monte Carlo computations.
In Section~\ref{sec:3}, we construct unbiased estimators for the EVPI and the EVPPI based on the MLMC method, and also briefly discuss some practical issues relating to implementation.
We conclude this paper with numerical experiments in Section~\ref{sec:4}.
\section{Expected value of information}\label{sec:2}
\subsection{Definitions}
Let $D$ be a finite set of decision options.
The task of a decision maker is to decide which option $d\in D$ is optimal under uncertainty of $X$.
Here $X$ is assumed to be a continuous random variable defined on the $s$-dimensional domain $\Omega_X \subseteq \RR^s$ with density $p_X$, and a monetary value function $f_d\colon \Omega_X \to \RR$ is assigned for each option $d\in D$.
Throughout this paper, we assume $f_d\in L^2(\Omega_X,p_X)$.
Under the risk neutrality assumption, the optimal option is one which maximizes the expected monetary value 
  \begin{align*}
   \EE_X\left[f_d\right]:=\int_{\Omega_X}f_d(x)p_X(x)\rd x.
  \end{align*}
Thus the expected monetary value without additional information is given by $\max_{d\in D}\EE_X\left[f_d\right]$.

\subsubsection{Expected value of perfect information}
Suppose that perfect information is available.
In this situation, a decision maker can decide an optimal option after eliminating uncertainty of $X$ completely.
Therefore, the monetary value for a decision maker after $X=x$ is indicated by perfect information is simply given by $\max_{d\in D}f_d(x)$.
As a result, the expected monetary value with perfect information becomes $\EE_X\left[\max_{d\in D}f_d \right]$.
The EVPI denotes how much the expected monetary value is increased by eliminating uncertainty of $X$.
Thus the EVPI is defined by
  \begin{align*}
   \EVPI := \EE_X\left[\max_{d\in D}f_d \right] - \max_{d\in D}\EE_X\left[f_d\right].
  \end{align*}
Note that the EVPI is equivalent to how much a decision maker is willing to pay for obtaining perfect information.
\subsubsection{Expected value of partial perfect information}
Assume that the random variable $X$ is separable into (possibly correlated) two random variables as $X=(X^{(1)},X^{(2)})$ with $\Omega_X = \Omega_{X^{(1)}}\times \Omega_{X^{(2)}}$, and that available information is perfect only for $X^{(1)}$.
In this situation, a decision maker can decide an optimal option under uncertainty of $X^{(2)}$ after eliminating uncertainty of $X^{(1)}$ completely.
Therefore, the monetary value for a decision maker after $X^{(1)}=x^{(1)}$ is indicated by partial perfect information is given by $\max_{d\in D}\EE_{X^{(2)}\mid X^{(1)}}\left[f_d(x^{(1)},\cdot)\right]$, where the expectation is taken with respect to $X^{(2)}$ given $X^{(1)}=x^{(1)}$.
As a result, the expected monetary value with partial perfect information for $X^{(1)}$ becomes $\EE_{X^{(1)}}\left[\max_{d\in D}\EE_{X^{(2)}\mid X^{(1)}}\left[f_d\right] \right]$.
Thus, similarly to the EVPI, the EVPPI on $X^{(1)}$ is defined by
  \begin{align*}
   \EVPPI_{X^{(1)}} := \EE_{X^{(1)}}\left[\max_{d\in D}\EE_{X^{(2)}\mid X^{(1)}}\left[f_d\right] \right] - \max_{d\in D}\EE_X\left[f_d\right].
  \end{align*}
Here we recall that the marginal density function of $X^{(1)}$ and the conditional density function of $X^{(2)}$ given $X^{(1)}=x^{(1)}$ are given by
  \begin{align*}
   p_{X^{(1)}}(x^{(1)}) = \int_{\Omega_{X^{(2)}}}p_{X}(x^{(1)},x^{(2)}) \rd x^{(2)},
  \end{align*}
and
  \begin{align*}
   p_{X^{(2)}\mid X^{(1)}}(x^{(2)}\mid x^{(1)}) = \frac{p_X(x^{(1)},x^{(2)})}{p_{X^{(1)}}(x^{(1)})},
  \end{align*}
respectively.

\subsection{Standard nested Monte Carlo}\label{subsec:nested_MC}
Since both the EVPI and the EVPPI are often difficult to calculate analytically, the Monte Carlo computations are used in practice.
Let us consider the EVPI first.
For $L,N\in \NN$, let $x_1,\ldots,x_N$ and $x'_1,\ldots,x'_L$ be i.i.d.\ random samples generated from $p_X$.
The EVPI is approximated by
  \begin{align}\label{eq:evpi_mc}
   \overline{\EVPI} = \frac{1}{N}\sum_{n=1}^{N}\max_{d\in D}f_d(x_n) - \max_{d\in D}\frac{1}{L}\sum_{l=1}^{L}f_d(x'_l).
  \end{align}
We can approximate the EVPPI in a similar way.
Let $L,M,N\in \NN$.
Let $x'_1,\ldots,x'_L$ be i.i.d.\ random samples generated from $p_X$, and $x^{(1)}_1,\ldots,x^{(1)}_N$ i.i.d.\ random samples generated from $p_{X^{(1)}}$.
For each $n$,  let $x^{(2)}_{1,n},\ldots,x^{(2)}_{M,n}$ be i.i.d.\ random samples generated from $p_{X^{(2)}\mid X^{(1)}}(\cdot\mid x^{(1)}_n)$.
Then the EVPPI is approximated by the following nested form
  \begin{align}\label{eq:evppi_mc}
   \overline{\EVPPI}_{X^{(1)}} = \frac{1}{N}\sum_{n=1}^{N}\max_{d\in D}\frac{1}{M}\sum_{m=1}^{M}f_d(x^{(1)}_n,x^{(2)}_{m,n}) - \max_{d\in D}\frac{1}{L}\sum_{l=1}^{L}f_d(x'_l).
  \end{align}
In the case where there is no correlation between $X^{(1)}$ and $X^{(2)}$, random samples $x^{(2)}_{1,n},\ldots,x^{(2)}_{M,n}$ used in the inner sum can be replaced by $M$ i.i.d.\ random samples generated from $p_{X^{(2)}}$ for all $n$.

Here we would emphasize that both the Monte Carlo estimators $\overline{\EVPI}$ and $\overline{\EVPPI}_{X^{(1)}}$ are biased.
That is, 
  \begin{align*}
   \EE\left[\overline{\EVPI}\right] \neq \EVPI \quad \text{and}\quad \EE\left[\overline{\EVPPI}_{X^{(1)}}\right] \neq \EVPPI_{X^{(1)}}.
  \end{align*}
This result follows directly from Jensen's inequality.
In case of the EVPI, we have
  \begin{align*}
   \EE\left[\overline{\EVPI}\right] & = \EE\left[\frac{1}{N}\sum_{n=1}^{N}\max_{d\in D}f_d(x_n)\right] - \EE\left[\max_{d\in D}\frac{1}{L}\sum_{l=1}^{L}f_d(x'_l)\right] \\
   & \leq \EE\left[\frac{1}{N}\sum_{n=1}^{N}\max_{d\in D}f_d(x_n)\right]  - \max_{d\in D}\EE\left[\frac{1}{L}\sum_{l=1}^{L}f_d(x'_l)\right] \\
   & = \EE_X\left[\max_{d\in D}f_d \right] - \max_{d\in D}\EE_X\left[f_d\right] = \EVPI,
  \end{align*}
where the inequality stems from Jensen's inequality.
It is clear that the first term of the EVPI can be estimated without any bias, whereas the second term is estimated with a positive bias.
We refer to \cite{MMW99} for a possible bounding technique to quantify the bias.
Therefore, in total, $\overline{\EVPI}$ is a downward biased estimator of $\EVPI$.
In case of the EVPPI, both the first and second terms of the EVPPI are estimated with positive biases.
Thus it is difficult to conclude whether the estimator $\overline{\EVPPI}_{X^{(1)}}$ is biased either upward or downward.
Nevertheless, these two biases are not cancelled out, so that we have $\EE\left[\overline{\EVPPI}_{X^{(1)}}\right] \neq \EVPPI_{X^{(1)}}$.

Moreover, there is a limitation in the asymptotic convergence rate of the estimator $\overline{\EVPPI}_{X^{(1)}}$.
For simplicity, let us focus on the first term of the EVPPI, which is approximated as
  \begin{align*}
   \EE_{X^{(1)}}\left[\max_{d\in D}\EE_{X^{(2)}\mid X^{(1)}}\left[f_d\right] \right] \sim \frac{1}{N}\sum_{n=1}^{N}\max_{d\in D}\frac{1}{M}\sum_{m=1}^{M}f_d(x^{(1)}_n,x^{(2)}_{m,n}).
  \end{align*}
It is clear from Jensen's inequality that the operator $\max_{d\in D}$ yields a positive bias.
Assuming that the bias depends on $X^{(1)}$ and decays uniformly at a certain rate $M^{-\gamma}$ for $\gamma>0$, it follows from the heuristic argument in \cite[Section~2]{AG16} that the approximation error itself decays at the rate
  \begin{align*}
  \max( N^{-1/2}, N^{-1/2}M^{-1/2}, M^{-\gamma}).
  \end{align*}
Notice that the total number of samples used is $C=MN$.
By putting $N=C^{w}$ and $M=C^{1-w}$ for $0<w<1$, the convergence rate of the standard nested Monte Carlo estimator in the total computational budget $C$ is given by the form $C^{-\rho}$, where
  \begin{align*}
  \rho = \min( w/2, 1/2, (1-w)\gamma).
  \end{align*}

Let us assume that the bias decays at a faster rate than the canonical Monte Carlo rate, i.e., $\gamma>1/2$.
When $w=1/2$, that is, when $N=M$, the resulting value for $\rho$ is $1/4$.
This convergence rate is also observed empirically in \cite{NGTS16}.
The best possible value for $\rho$ is actually $\gamma/(1+2\gamma)$, which is strictly less than $1/2$ and is attained when $w=2\gamma/(1+2\gamma)$.
This result suggests fewer samples for the inner sum and more for the outer sum, which is opposite from what is empirically recommended in \cite{BKOC07}.
Nevertheless, the above argument concludes that the square-root convergence rate in $C$ cannot be achieved by the standard nested Monte Carlo estimator.

\section{Construction of unbiased estimators}\label{sec:3}
We now move on to constructing unbiased Monte Carlo estimators for the EVPI and the EVPPI based on the MLMC method.
\subsection{Expected value of perfect information}
We first construct unbiased Monte Carlo estimators for the EVPI.
For $N\in \NN$, let us denote
  \begin{align*}
   Q(N)=\max_{d\in D}\frac{1}{N}\sum_{n=1}^{N}f_d(x_n),
  \end{align*}
where $x_1,\ldots,x_N$ are i.i.d.\ random samples generated from $p_X$.
Then the following simple but interesting properties hold:
  \begin{align*}
   \EE\left[Q(1)\right] = \EE\left[\max_{d\in D}f_d(x_1)\right] = \EE_X\left[\max_{d\in D}f_d \right] ,
  \end{align*}
and
  \begin{align*}
   \lim_{N\to \infty}Q(N) = \max_{d\in D}\lim_{N\to \infty}\frac{1}{N}\sum_{n=1}^{N}f_d(x_n) = \max_{d\in D}\EE_X\left[f_d\right] ,
  \end{align*}
where the latter property stems from the law of large numbers.
Hence it is trivial that
  \begin{align*}
   \lim_{N\to \infty}\EE\left[Q(N)\right] = \max_{d\in D}\EE_X\left[f_d\right] .
  \end{align*}
Therefore, the EVPI can be rewritten into
  \begin{align}\label{eq:evpi_rewrite}
   \EVPI = \EE\left[Q(1)\right] - \lim_{N\to \infty}\EE\left[Q(N)\right].
  \end{align}
Based on this finding and the idea of the MLMC method, unbiased Monte Carlo estimators for the EVPI can be constructed.
For a fixed integer $b\geq 2$, we have the telescopic representation for the right-hand side of (\ref{eq:evpi_rewrite})
  \begin{align*}
   \EE\left[Q(1)\right] - \lim_{N\to \infty}\EE\left[Q(N)\right] 
   & = \left\{ \EE\left[Q(1)\right]-\EE\left[Q(b)\right]\right\}+\left\{ \EE\left[Q(b)\right]-\EE\left[Q(b^2)\right]\right\}+\cdots \\
   & = \sum_{l=1}^{\infty}\left\{ \EE\left[Q(b^{l-1})\right]-\EE\left[Q(b^l)\right]\right\} \\
   & = \sum_{l=1}^{\infty}\EE\left[Q(b^{l-1})-Q(b^l)\right],
  \end{align*}
where the last equality stems from the linearity of expectation.

Now let us introduce a positive integer-valued independent random variable $L$ with a probability mass function $p_L$ such that $p_L(l)>0$ for all $l\in \NN$.
Then the first \emph{single term estimator} of the EVPI is given by
  \begin{align}\label{eq:est_EVPI_single}
    \frac{1}{N}\sum_{n=1}^{N}Y_{l_n}^{\single},
  \end{align}
where $l_1,\ldots,l_N$ are i.i.d.\ random samples generated from $p_L$, and for each $l\in \NN$ we define
  \begin{align*}
    Y_{l}^{\single} = \frac{1}{p_L(l)}\left[\frac{1}{b}\sum_{\lambda=0}^{b-1}\max_{d\in D}\frac{1}{b^{l-1}}\sum_{m=1}^{b^{l-1}}f_d(x_{\lambda b^{l-1}+m})-\max_{d\in D}\frac{1}{b^{l}}\sum_{m=1}^{b^{l}}f_d(x_m)\right] ,
  \end{align*}
where $x_1,\ldots,x_{b^{l}}$ are i.i.d.\ random samples generated from $p_X$.
Note that the same samples $x_1,\ldots,x_{b^{l}}$ are commonly used in the first and second terms in $Y_{l}^{\single}$.
It can be seen that this estimator is unbiased:
  \begin{align*}
   \EE\left[\frac{1}{N}\sum_{n=1}^{N}Y_{l_n}^{\single} \right] & = \EE\left[\sum_{l=1}^{\infty}p_L(l)Y_{l}^{\single} \right] = \sum_{l=1}^{\infty}p_L(l)\EE\left[Y_{l}^{\single} \right] \\
   & = \sum_{l=1}^{\infty}\EE\left[Q(b^{l-1})-Q(b^l)\right] = \EVPI.
  \end{align*}

The second \emph{coupled sum estimator} of the EVPI is given by
  \begin{align}\label{eq:est_EVPI_coupled}
    \frac{1}{N}\sum_{n=1}^{N}Y_{l_n}^{\coupled},
  \end{align}
where $l_1,\ldots,l_N$ are i.i.d.\ random samples generated from $p_L$, and for each $l\in \NN$ we define
  \begin{align*}
    Y_{l}^{\coupled} = \sum_{j=1}^{l}\frac{1}{\sum_{k=j}^{\infty}p_L(k)}& \left[\frac{1}{b^{l-j+1}}\sum_{\lambda=0}^{b^{l-j+1}-1}\max_{d\in D}\frac{1}{b^{j-1}}\sum_{m=1}^{b^{j-1}}f_d(x_{\lambda b^{j-1}+m}) \right. \\
   & \qquad - \left. \frac{1}{b^{l-j}}\sum_{\lambda=0}^{b^{l-j}-1}\max_{d\in D}\frac{1}{b^j}\sum_{m=1}^{b^j}f_d(x_{\lambda b^{j}+m})\right] ,
  \end{align*}
where $x_1,\ldots,x_{b^{l}}$ are i.i.d.\ random samples generated from $p_X$.
Note that the same samples $x_1,\ldots,x_{b^{l}}$ are commonly used in every term in $Y_{l}^{\coupled}$.
It can be seen that this estimator is also unbiased:
  \begin{align*}
   \EE\left[\frac{1}{N}\sum_{n=1}^{N}Y_{l_n}^{\coupled} \right] & = \EE\left[\sum_{l=1}^{\infty}p_L(l)Y_{l}^{\coupled} \right] = \sum_{l=1}^{\infty}p_L(l)\EE\left[Y_{l}^{\coupled} \right] \\
   & = \sum_{l=1}^{\infty}\sum_{j=1}^{l}\frac{p_L(l)}{\sum_{k=j}^{\infty}p_L(k)}\EE\left[Q(b^{j-1})-Q(b^j) \right] \\
   & = \sum_{j=1}^{\infty}\sum_{l=j}^{\infty}\frac{p_L(l)}{\sum_{k=j}^{\infty}p_L(k)}\EE\left[Q(b^{j-1})-Q(b^j) \right] \\
   & = \sum_{j=1}^{\infty}\EE\left[Q(b^{j-1})-Q(b^j)\right] = \EVPI ,
  \end{align*}
where the fourth equality is given by swapping the order of sums.
\subsection{Expected value of partial perfect information}
In a way similar to that for the EVPI, it is possible to construct unbiased Monte Carlo estimators for the EVPPI.
Since unbiased estimators for the EVPI have been constructed already, it suffices to construct unbiased estimators for
  \begin{align*}
   \EVPI - \EVPPI_{X^{(1)}} = \EE_X\left[\max_{d\in D}f_d \right] - \EE_{X^{(1)}}\left[\max_{d\in D}\EE_{X^{(2)}\mid X^{(1)}}\left[f_d\right] \right].
  \end{align*}
For $N\in \NN$ and $x^{(1)}\in \Omega_{X^{(1)}}$, let us denote
  \begin{align*}
   Q(N;x^{(1)}) = \max_{d\in D}\frac{1}{N}\sum_{n=1}^{N}f_d(x^{(1)},x^{(2)}_n) ,
  \end{align*}
where $x^{(2)}_1,\ldots,x^{(2)}_N$ are generated independently and randomly from $p_{X^{(2)}\mid X^{(1)}}(\cdot\mid x^{(1)})$.
Then the following properties hold:
  \begin{align*}
   \EE\left[Q(1;x^{(1)})\right] = \EE_{X^{(2)}\mid X^{(1)}}\left[\max_{d\in D}f_d(x^{(1)},\cdot) \right] ,
  \end{align*}
and
  \begin{align*}
   \lim_{N\to \infty}\EE\left[Q(N;x^{(1)})\right] =\max_{d\in D}\EE_{X^{(2)}\mid X^{(1)}}\left[f_d(x^{(1)},\cdot)\right] ,
  \end{align*}
where the latter property stems again from the law of large numbers.
Using these results, we have
  \begin{align*}
   \EVPI - \EVPPI_{X^{(1)}} & =  \EE_{X^{(1)}}\EE_{X^{(2)}\mid X^{(1)}}\left[\max_{d\in D}f_d \right] - \EE_{X^{(1)}}\left[\max_{d\in D}\EE_{X^{(2)}\mid X^{(1)}}\left[f_d\right] \right] \\
   & =  \EE_{X^{(1)}}\EE\left[Q(1;\cdot)\right] - \EE_{X^{(1)}}\left[\lim_{N\to \infty}\EE\left[Q(N;\cdot)\right]\right] \\
   & = \EE_{X^{(1)}}\left[\EE\left[Q(1;\cdot)\right] - \lim_{N\to \infty}\EE\left[Q(N;\cdot)\right]\right] .
  \end{align*}
For a fixed integer $b\geq 2$, we have the telescopic representation
  \begin{align*}
   \EE_{X^{(1)}}\left[\EE\left[Q(1;\cdot)\right] - \lim_{N\to \infty}\EE\left[Q(N;\cdot)\right]\right] & = \EE_{X^{(1)}}\left[ \sum_{l=1}^{\infty}\left\{ \EE\left[Q(b^{l-1};\cdot)\right]-\EE\left[Q(b^l;\cdot)\right]\right\}\right] \\
   & = \EE_{X^{(1)}}\left[ \sum_{l=1}^{\infty}\EE\left[Q(b^{l-1};\cdot)-Q(b^l;\cdot)\right] \right].
  \end{align*}

Now let us introduce a positive integer-valued independent random variable $J$ with a probability mass function $p_J$ such that $p_J(j)>0$ for all $j\in \NN$.
Then the first \emph{single term estimator} of $\EVPI - \EVPPI_{X^{(1)}}$ is given by
  \begin{align}\label{eq:est_EVPPI_single}
    \frac{1}{N}\sum_{n=1}^{N}Z_{l_n, x^{(1)}_n}^{\single},
  \end{align}
where $l_1,\ldots,l_N$ and $x^{(1)}_1,\ldots,x^{(1)}_N$ are i.i.d.\ random samples generated from $p_L$ and $p_{X^{(1)}}$, respectively, and for each $l\in \NN$ and $x^{(1)}\in \Omega_{X^{(1)}}$ we define
  \begin{align*}
    Z_{l, x^{(1)}}^{\single} = \frac{1}{p_L(l)}\left[\frac{1}{b}\sum_{\lambda=0}^{b-1}\max_{d\in D}\frac{1}{b^{l-1}}\sum_{m=1}^{b^{l-1}}f_d(x^{(1)},x^{(2)}_{\lambda b^{l-1}+m})-\max_{d\in D}\frac{1}{b^{l}}\sum_{m=1}^{b^{l}}f_d(x^{(1)},x^{(2)}_m)\right] ,
  \end{align*}
where $x^{(2)}_1,\ldots,x^{(2)}_{b^l}$ are i.i.d.\ random samples generated from $p_{X^{(2)}\mid X^{(1)}}(\cdot\mid x^{(1)})$.
It can be seen that this estimator is unbiased:
  \begin{align*}
   \EE\left[\frac{1}{N}\sum_{n=1}^{N}Z_{l_n, x^{(1)}_n}^{\single} \right] & = \EE\left[\sum_{l=1}^{\infty}p_L(l)\EE_{X^{(1)}}\left[Z_{l_n, \cdot}^{\single}\right]  \right] = \EE_{X^{(1)}}\left[\sum_{l=1}^{\infty}p_L(l)\EE\left[Z_{l_n, \cdot}^{\single} \right]\right] \\
   & = \EE_{X^{(1)}}\left[\sum_{l=1}^{\infty}\EE\left[Q(b^{l-1};\cdot)-Q(b^l;\cdot)\right]\right] = \EVPI.
  \end{align*}

The second \emph{coupled sum estimator} of $\EVPI - \EVPPI_{X^{(1)}}$ is given by
  \begin{align}\label{eq:est_EVPPI_coupled}
    \frac{1}{N}\sum_{n=1}^{N}Z_{l_n, x^{(1)}_n}^{\coupled},
  \end{align}
where $l_1,\ldots,l_N$ and $x^{(1)}_1,\ldots,x^{(1)}_N$ are i.i.d.\ random samples generated from $p_L$ and $p_{X^{(1)}}$, respectively, and for each $l\in \NN$ and $x^{(1)}\in \Omega_{X^{(1)}}$ we define
  \begin{align*}
    Z_{l, x^{(1)}}^{\coupled} = \sum_{j=1}^{l}\frac{1}{\sum_{k=j}^{\infty}p_L(k)}& \left[\frac{1}{b^{l-j+1}}\sum_{\lambda=0}^{b^{l-j+1}-1}\max_{d\in D}\frac{1}{b^{j-1}}\sum_{m=1}^{b^{j-1}}f_d(x^{(1)},x^{(2)}_{\lambda b^{j-1}+m}) \right. \\
   & \qquad - \left. \frac{1}{b^{l-j}}\sum_{\lambda=0}^{b^{l-j}-1}\max_{d\in D}\frac{1}{b^j}\sum_{m=1}^{b^j}f_d(x^{(1)},x^{(2)}_{\lambda b^{j}+m})\right] ,
  \end{align*}
where $x^{(2)}_1,\ldots,x^{(2)}_{b^l}$ are i.i.d.\ random samples generated from $p_{X^{(2)}\mid X^{(1)}}(\cdot\mid x^{(1)})$.
Note that the same samples $x_1,\ldots,x_{b^{l}}$ are commonly used in every term in $Z_{l, x^{(1)}}^{\coupled}$.
This estimator can be shown unbiased in a similar way as above.

Therefore, $\EVPPI_{X^{(1)}}$ can be estimated without any bias by
  \begin{align*}
   \frac{1}{N}\sum_{n=1}^{N}(Y_{l_n}^{\dagger}-Z_{l'_n, x^{(1)}_n}^{\ddagger}),
  \end{align*}
where $\dagger,\ddagger\in \{\single, \coupled\}$, and $l_1,\ldots,l_N$ and $l'_1,\ldots,l'_N$ are i.i.d.\ random samples generated from $p_L$.
Note that even if one randomly chosen level $l_n$ is used commonly for each $n$ as
  \begin{align}\label{eq:evppi_mlmc}
   \frac{1}{N}\sum_{n=1}^{N}(Y_{l_n}^{\dagger}-Z_{l_n, x^{(1)}_n}^{\ddagger}),
  \end{align}
the resulting estimator is still unbiased.

\subsection{Implementation}\label{subsec:imple}
So far, we do not specify the probability mass function $p_J$ in both the estimators for the EVPI and the EVPPI.
It should be chosen so as for both the variance of the estimator and the expected computational cost to be finite \cite{Giles15}.

Let us consider the estimator (\ref{eq:est_EVPI_single}) of the EVPI as an example.
The variance and the expected computational cost per one sample of (\ref{eq:est_EVPI_single}) are given by
  \begin{align*}
   \sum_{l=1}^{\infty}\frac{\EE\left[(Y_l^{\single})^2\right]}{p_L(l)}\quad \text{and} \quad \sum_{l=1}^{\infty}p_L(l)b^l,
  \end{align*}
respectively, where the expectation is taken with respect to random samples $x_1,\ldots,x_{b^l}$ for every $l\in \NN$.
It can be easily shown that the optimal choice for $p_L$, which minimizes 
  \begin{align*}
   \sum_{l=1}^{\infty}\left( \frac{\EE\left[(Y_l^{\single})^2\right]}{p_L(l)} + \lambda p_L(l)b^l\right),
  \end{align*}
for some Lagrange multiplier $\lambda$, is given by
  \begin{align*}
   p_L(l) = \sqrt{\frac{\EE\left[(Y_l^{\single})^2\right]}{b^l}}\left( \sum_{j=1}^{\infty}\sqrt{\frac{\EE\left[(Y_j^{\single})^2\right]}{b^j}}\right)^{-1},
  \end{align*}
if the sum over $j$ is finite.
In order to achieve the root mean square error less than $\varepsilon$, the required expected total computational cost is 
  \begin{align*}
   C = \varepsilon^{-2}\left( \sum_{j=1}^{\infty}\sqrt{\EE\left[(Y_j^{\single})^2\right] b^j}\right)^{2} ,
  \end{align*}
again if the sum over $j$ is finite, see \cite[Section~2.2]{Giles15}.
Note that the above argument also holds for other estimators (\ref{eq:est_EVPI_coupled}), (\ref{eq:est_EVPPI_single}) and (\ref{eq:est_EVPPI_coupled}).
Therefore, in preferable cases, our unbiased estimators can achieve the square-root convergence rate in $C$, which is not possible for the nested Monte Carlo computation as heuristically discussed in the last section.
Note that if the sum over $j$ is not finite, on the other hand, there is no theoretical foundation on the convergence behavior of our estimators, so that one may observe a deteriorated convergence rate in practical computations.

In practice, we need a more reasonable choice for $p_L$ in some sense since the expectations $\EE\left[(Y_l^{\single})^2\right]$ are not known in advance.
As an alternative approach, let us specify the form of $p_L$ as $p_L(l)=(1-r)r^{l-1}$ for all $l\geq 1$ with $0<r<1$.
Let us assume that the expectations $\EE\left[(Y_l^{\single})^2\right]$ decay at a rate $b^{-2q l}$ for some $q>1/2$.
In order for both the variance of the estimator, in which $\EE\left[(Y_l^{\single})^2\right]$ is replaced by $b^{-2q l}$, and the expected computational cost to be finite, it suffices that $b^{-2q}<r<b^{-1}$ holds.
Moreover, the optimal choice for $p_L$ obtained in the last paragraph gives $r=b^{-(2q+1)/2}$.
The same value of $r$ can be obtained, as done in \cite{BG15} where the special case with $b=2$ and $q=1$ is considered, by minimizing the work-normalized variance
  \begin{align*}
   \left(\sum_{l=1}^{\infty}\frac{b^{-2q l}}{p_L(l)}\right)\left(\sum_{l=1}^{\infty}p_L(l)b^l \right) .
  \end{align*}

Furthermore, in practical applications, one may set the total computational cost $C$ instead of $N$, i.e., the number of i.i.d.\ copies used in the estimators.
In this case, we first generate a random sequence $l_1,l_2,\ldots \in \NN$ independently from $p_L$ and then define $N$ by \[N:=\max\{j \colon b^{l_1}+b^{l_2}+\cdots + b^{l_j}\leq C\}.\]

\section{Numerical experiments}\label{sec:4}
Finally, we conduct numerical experiments for a simple toy problem.
In order to evaluate the approximation error quantitatively, we design a toy problem such that the EVPI and the EVPPI can be calculated analytically.

\subsection{Toy problem}
Let us consider the following setting.
Let $D=\{d_1,d_2\}$ be a set of two possible actions which can be taken by a decision maker under uncertainty of $X=(X_j)_{j=1,\ldots,s}$.
For $x=(x_1,\ldots,x_s)\in \RR^s$ we define
\begin{align*}
f_{d_1}(x):= w_0+\sum_{j=1}^{s}w_jx_j\quad \text{and}\quad f_{d_2}(x):= 0 ,
\end{align*}
where $w_0\in \RR$ and $w_1,\ldots,w_s\in \RR\setminus \{0\}$.
The prior probability density of $X$ is given by
\begin{align*}
p_X(x) = \prod_{j=1}^{s}p_{X_j}(x_j) = \prod_{j=1}^{s}\frac{1}{\sqrt{2\pi}\sigma_j}\exp\left\{-\frac{(x_j-\mu_j)^2}{2\sigma_j^2}\right\},
\end{align*}
for given $\mu_1,\ldots,\mu_s\in \RR$ and $\sigma_1,\ldots,\sigma_s>0$. Then we have
\begin{align*}
\max_{d\in \{d_1,d_2\}}\EE_X\left[f_d\right] := \max\left\{ \int_{\RR^s}f_{d_1}(x)p(x)\rd x, 0\right\} = \max\left\{ w_0+\sum_{j=1}^{s}w_j\mu_j, 0\right\}.
\end{align*}

Now let $u$ be a subset of $\{1,\ldots,s\}$. 
For simplicity, let us focus on the case $u=\{1,\ldots,|u|\}$. 
We write $-u=\{1,\ldots,s\}\setminus u$, $X_u=(X_j)_{j\in u}$ and $X_{-u}=(X_j)_{j\in -u}$.
The EVPPI on $X_u$ can be calculated analytically as follows.
Since there is no correlation between $X_u$ and $X_{-u}$, we have
\begin{align*}
& \EE_{X_u}\left[\max_{d\in \{d_1,d_2\}}\EE_{X_{-u}}\left[f_d\right] \right] \\
& \quad = \int_{\RR^{|u|}}\max\left\{ \int_{\RR^{s-|u|}}f_{d_1}(x_u,x_{-u})\prod_{j\in -u}p_{X_j}(x_j)\rd x_{-u}, 0\right\}\prod_{j\in u}p_{X_j}(x_j)\rd x_u \\
& \quad = \int_{\RR^{|u|}}\max\left\{ \sum_{j\in u}w_jx_j+w_0+\sum_{j\in -u}w_j\mu_j, 0\right\}\prod_{j\in u}p_{X_j}(x_j)\rd x_u \\
& \quad = \int_{\Omega_{\geq 0}}\left( \sum_{j\in u}w_jx_j+w_0+\sum_{j\in -u}w_j\mu_j\right) \prod_{j\in u}p_{X_j}(x_j)\rd x_u ,
\end{align*}
where we write $\Omega_{\geq 0}=\{x_u\in \RR^{|u|}\colon \sum_{j\in u}w_jx_j+w_0+\sum_{j\in -u}w_j\mu_j\geq 0\}$.
By changing the variables according to $y= \bsA x_u$ where
\begin{align*}
\bsA = \left[ \begin{array}{cccc}
 1 & 0 & \cdots & 0 \\
 0 & 1 & \cdots & 0 \\
 \vdots & \vdots & \ddots & \vdots \\
 w_1 & w_2 & \cdots & w_{|u|} 
\end{array} \right] \in \RR^{|u|\times |u|} ,
\end{align*}
we have
\begin{align*}
\sum_{j\in u}w_jx_j = y_{|u|} ,
\end{align*}
i.e., the sum of $x_j$'s is nothing but the last component of $y$.
Moreover, the probability density of $y_{|u|}$ is the normal density with the mean $\sum_{j\in u}w_j\mu_j$ and the variance $\sum_{j\in u}(w_j\sigma_j)^2$, the above integral can be written into
\begin{align*}
\EE_{X_u}\left[\max_{d\in \{d_1,d_2\}}\EE_{X_{-u}}\left[f_d\right] \right] & = \int_{\Omega_{\geq 0}}\left( \sum_{j\in u}w_jx_j+w_0+\sum_{j\in -u}w_j\mu_j\right) \prod_{j\in u}p_{X_j}(x_j)\rd x_u \\
& = \int_{a}^{\infty}\left( y_{|u|}-a\right) p(y_{|u|})\rd y_{|u|} ,
\end{align*}
where we write $$a=-w_0-\sum_{j\in -u}w_j\mu_j.$$ Finally, the last integral equals
\begin{align*}
\EE_{X_u}\left[\max_{d\in \{d_1,d_2\}}\EE_{X_{-u}}\left[f_d\right] \right] = \left[ 1-\Phi\left( -\frac{\mu_{\all}}{\sigma_u}\right)\right]\mu_{\all} + \phi\left( -\frac{\mu_{\all}}{\sigma_u}\right) \sigma_u,
\end{align*}
where we write
\begin{align*}
\mu_{\all} = w_0+\sum_{j=1}^{s}w_j\mu_j ,
\end{align*}
and
\begin{align*}
\sigma_u^2 = \sum_{j\in u}(w_j\sigma_j)^2 .
\end{align*}
Further, $\phi$ denotes the standard normal density function and $\Phi$ does the cumulative distribution function for $\phi$.
Thus, the EVPPI on $X_u$ is given by
\begin{align*}
\EVPPI_{X_u} = \left[ 1-\Phi\left( -\frac{\mu_{\all}}{\sigma_u}\right)\right]\mu_{\all} + \phi\left( -\frac{\mu_{\all}}{\sigma_u}\right) \sigma_u -\max\left\{ \mu_{\all}, 0\right\}.
\end{align*}
Note that the analytical expression for the EVPI is given by setting $u=\{1,\ldots,s\}$.

\subsection{Numerical results}
In what follows, we focus on the case where $s=5$ and $\omega_j=1$, $\mu_j=0$, $\sigma_j=1$ for all $j=1,\ldots,s$ for the sake of simplicity.
First, let us consider the EVPI computation.
The analytical calculation of the EVPI is given by $0.892\ldots$.
We use the three estimators (\ref{eq:evpi_mc}), (\ref{eq:est_EVPI_single}) and (\ref{eq:est_EVPI_coupled}) of the EVPI for approximate evaluations.
When the total computational budget equals $C$, the naive Monte Carlo estimator (\ref{eq:evpi_mc}) is set by $L=N=C$, whereas our proposed estimators (\ref{eq:est_EVPI_single}) and (\ref{eq:est_EVPI_coupled}) are set as described in the last paragraph of Subsection~\ref{subsec:imple} with $b=2$ and $p_L(l)=(1-r)r^{l-1}$ where $r=2^{-3/2}$.
For a given total computational budget, 100 independent computations are conducted for each estimator.

\begin{figure}
\begin{center}
\includegraphics[width=0.32\textwidth]{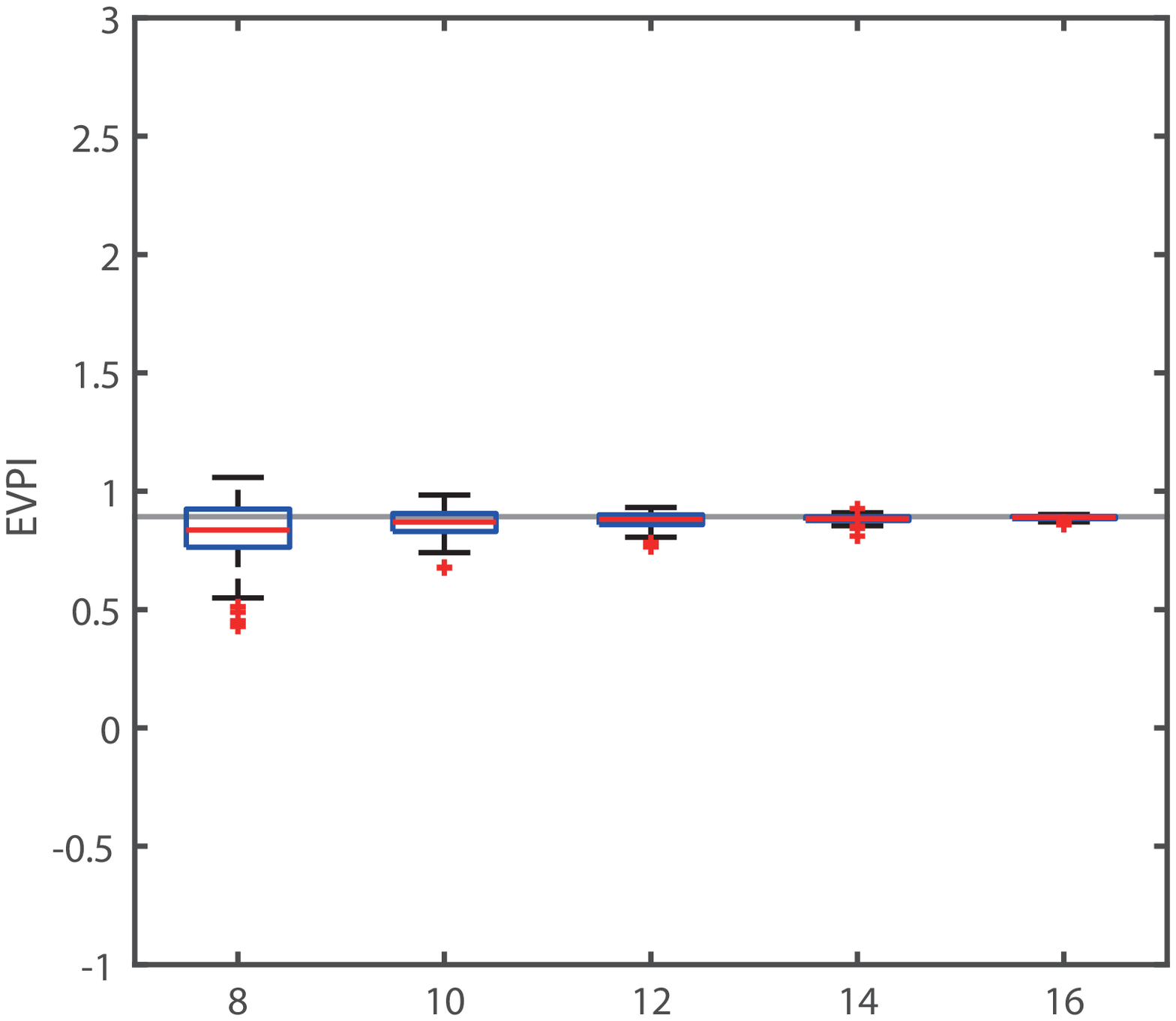}
\includegraphics[width=0.32\textwidth]{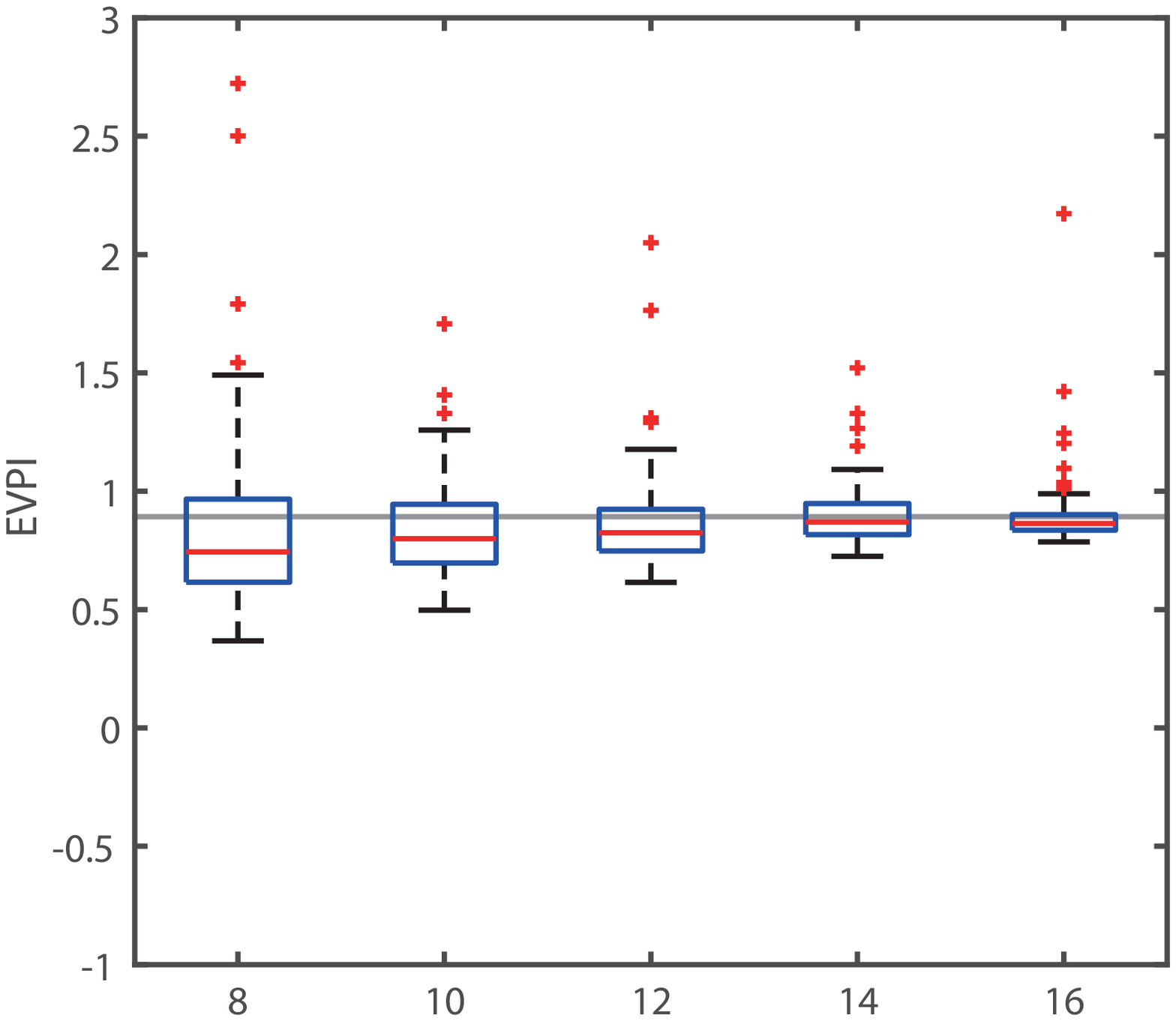}
\includegraphics[width=0.32\textwidth]{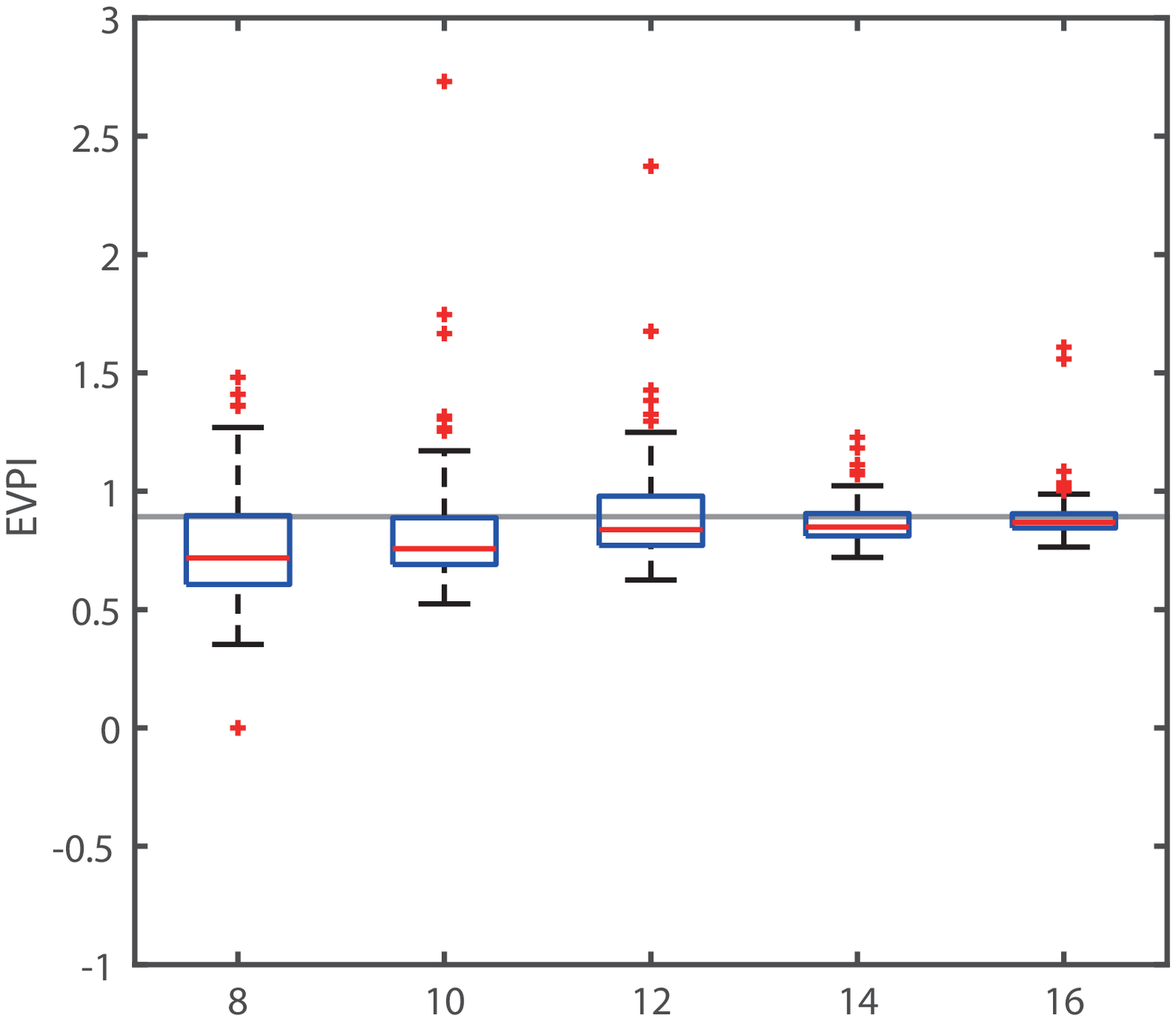}
\caption{Boxplots for 100 independent EVPI computations by the naive Monte Carlo estimator (left), the single term estimator (middle), and the coupled sum estimator (right) with the total computational budgets $C=2^8,\ldots,2^{16}$.}
\label{fig:evpi}
\end{center}
\end{figure}

Figure~\ref{fig:evpi} compares the boxplots of the EVPI computations obtained by three estimators as functions of $C=2^m$ with $m=8,10,\ldots,16$.
It can be seen that the naive Monte Carlo estimator gives more accurate results than our estimators.
In case of the EVPI, the naive Monte Carlo estimator is not of the nested form so that the approximation error decays at a rate of $C^{-1/2}$ if the bias decays at a faster rate than the canonical Monte Carlo rate.
Moreover, it can be expected that the variances of our estimators are much larger than that of the naive Monte Carlo estimator, which yields wider variations among the independent EVPI computations by our estimators as well as the difficulty in confirming the unbiasedness of our estimators when the total computational budget is small.

Let us move on to the EVPPI computation, in which case the situation changes significantly.
Because of the invariance of parameters, we focus on computing the EVPPI's on $X^{(1)}=X_1,(X_1,X_2), (X_1,X_2,X_3),(X_1,X_2,X_3,X_4)$.
The analytical calculations of the EVPPI's are given by $0.389\ldots$, $0.564\ldots$, $0.690\ldots$, and $0.797\ldots$, respectively.
We use the three estimators (\ref{eq:evppi_mc}), (\ref{eq:evppi_mlmc}) with the single term estimator ($\dagger,\ddagger=\single$), and (\ref{eq:evppi_mlmc}) with the coupled sum estimator ($\dagger,\ddagger=\coupled$) of the EVPI for approximate evaluations.
When the total computational budget equals $C$, the nested Monte Carlo estimator (\ref{eq:evppi_mc}) is set by $L=C$, $M=\lfloor C^{1/3}\rfloor$ and $N=\lfloor C^{2/3}\rfloor$ as heuristically suggested in Subsection~\ref{subsec:nested_MC} with $\gamma=1$.
In fact, although we also conducted the same numerical experiments by setting both $(M,N)=(\lfloor C^{1/2}\rfloor, \lfloor C^{1/2}\rfloor)$ and $(M,N)=(\lfloor C^{2/3}\rfloor, \lfloor C^{1/3}\rfloor)$, we obtained similar results to those with $(M,N)=(\lfloor C^{1/3}\rfloor, \lfloor C^{2/3}\rfloor)$, which are thus omitted in this paper.
Our proposed estimators (\ref{eq:evppi_mlmc}) with $\dagger,\ddagger\in \{\single, \coupled\}$ are set as described in the last paragraph of Subsection~\ref{subsec:imple} with $b=2$ and $p_L(l)=(1-r)r^{l-1}$ where $r=2^{-3/2}$.
For a given total computational budget, 100 independent computations are conducted for each estimator.

\begin{figure}
\begin{center}
\includegraphics[width=0.32\textwidth]{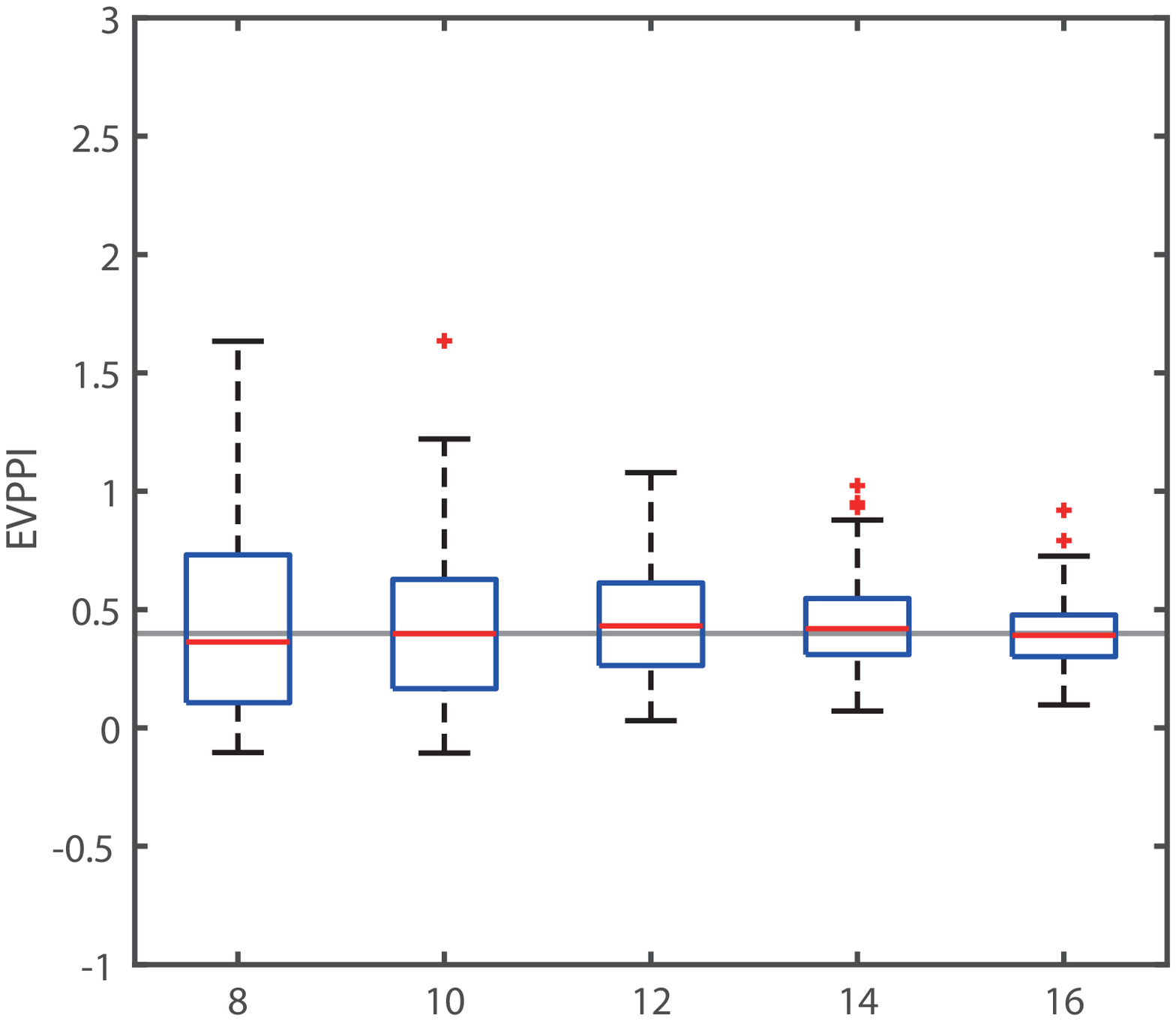}
\includegraphics[width=0.32\textwidth]{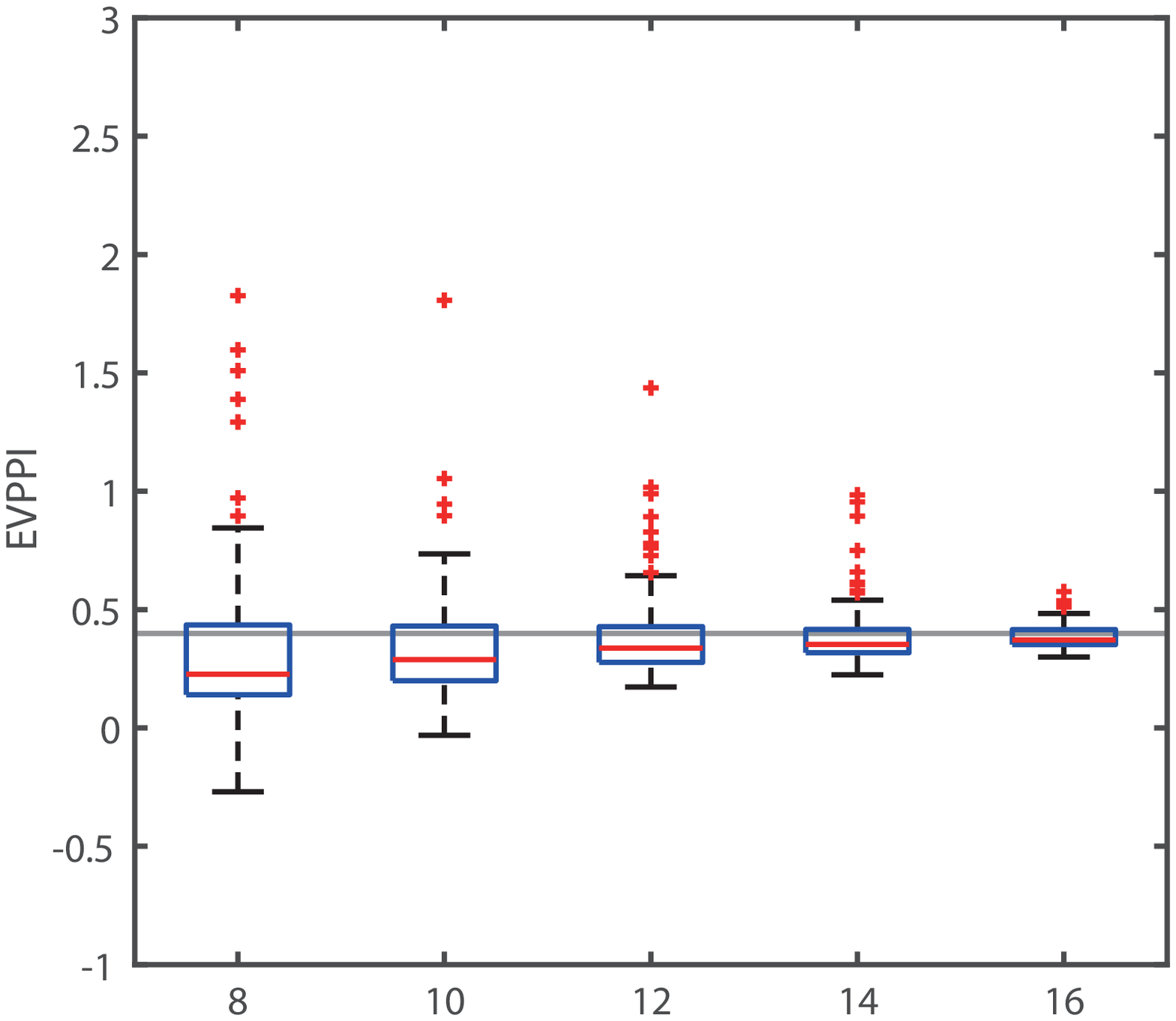}
\includegraphics[width=0.32\textwidth]{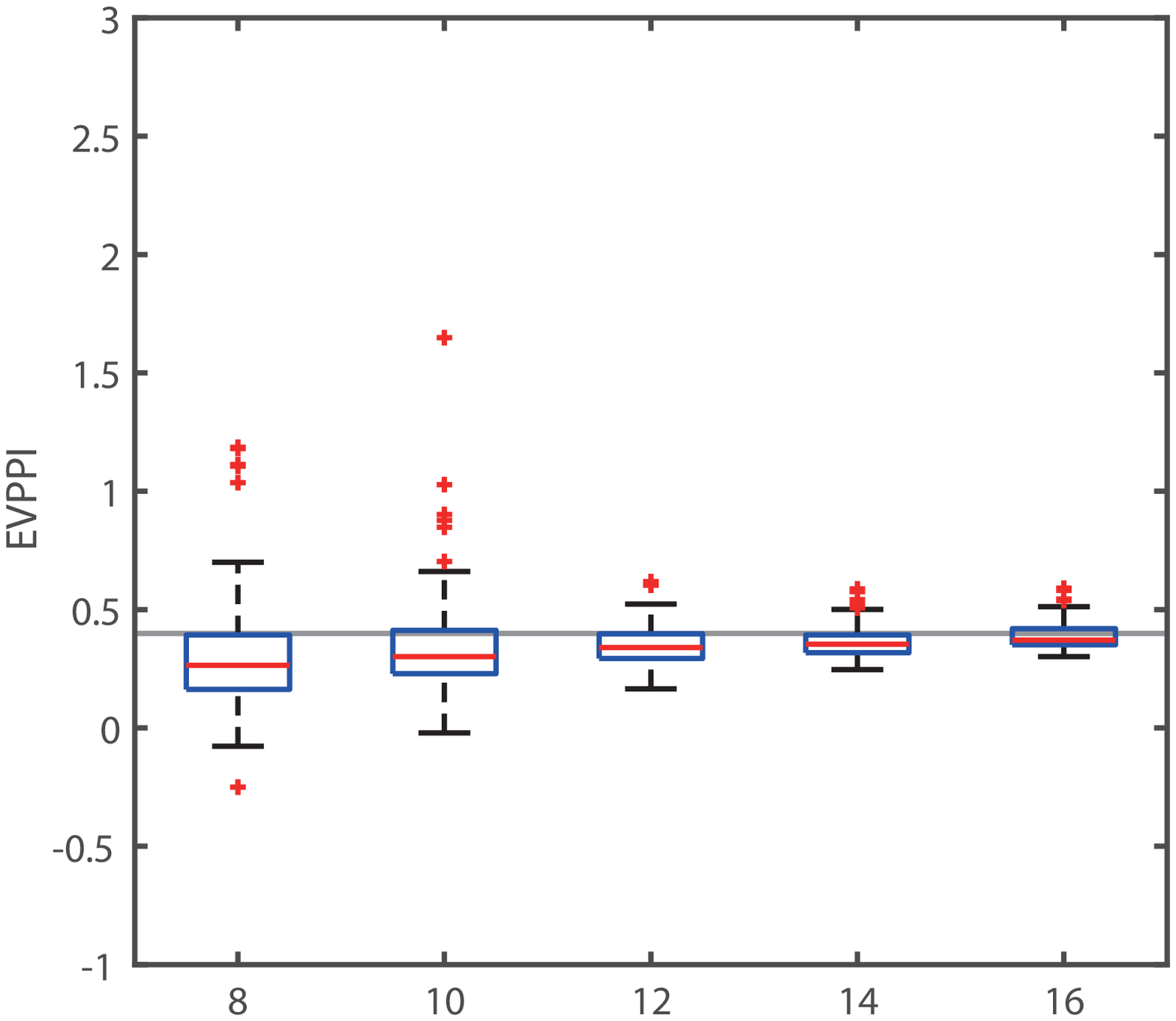}\\
\includegraphics[width=0.32\textwidth]{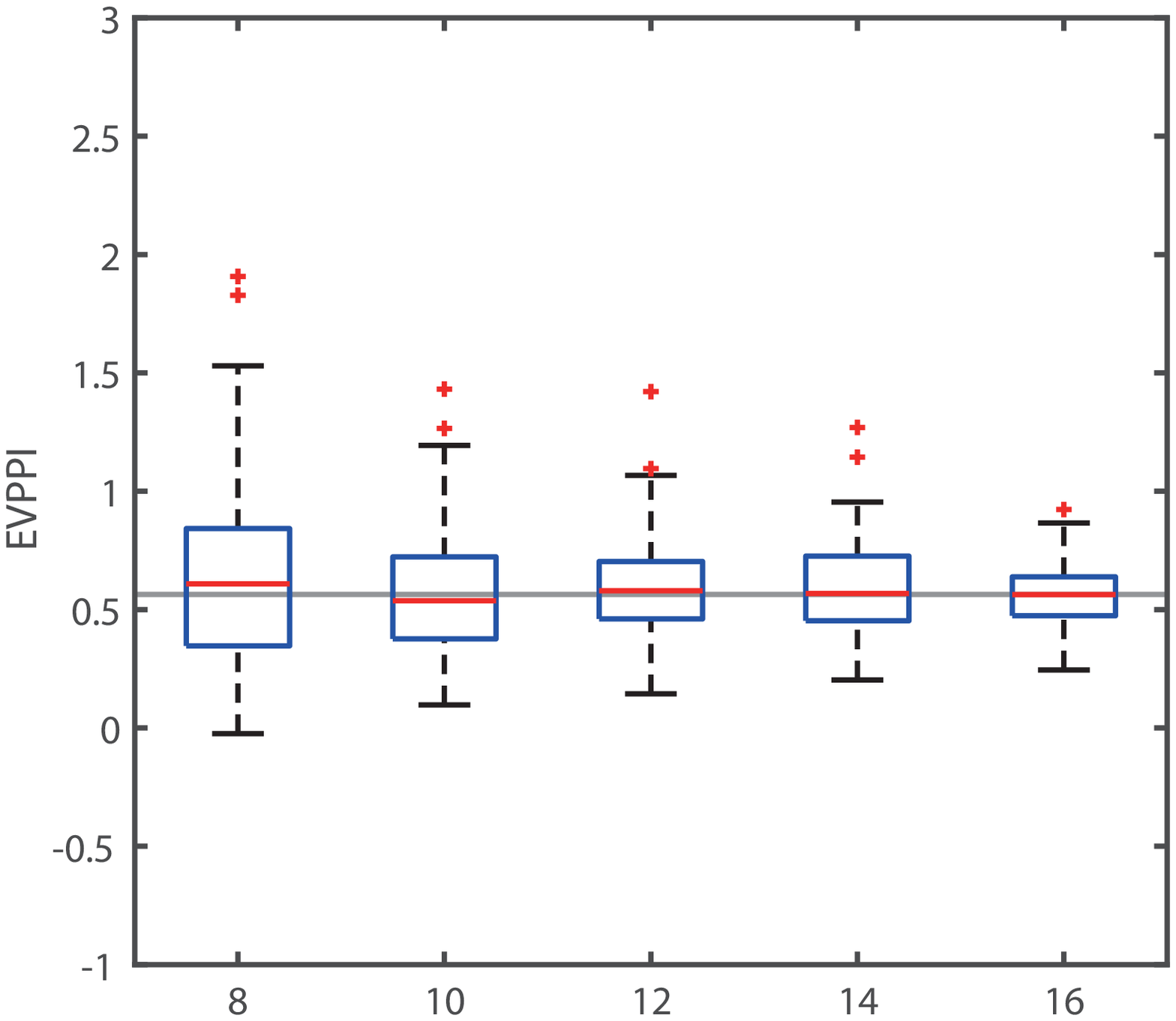}
\includegraphics[width=0.32\textwidth]{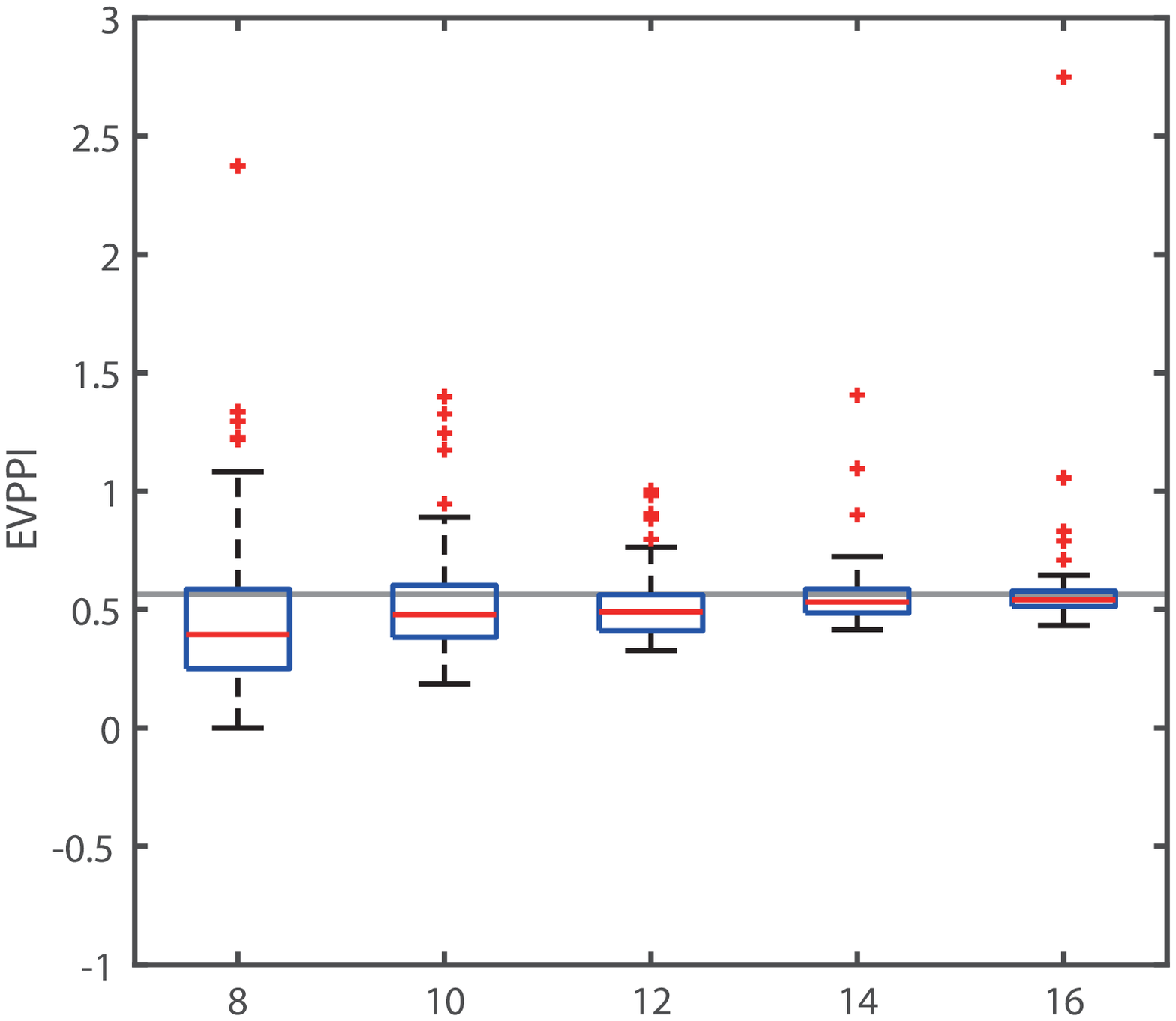}
\includegraphics[width=0.32\textwidth]{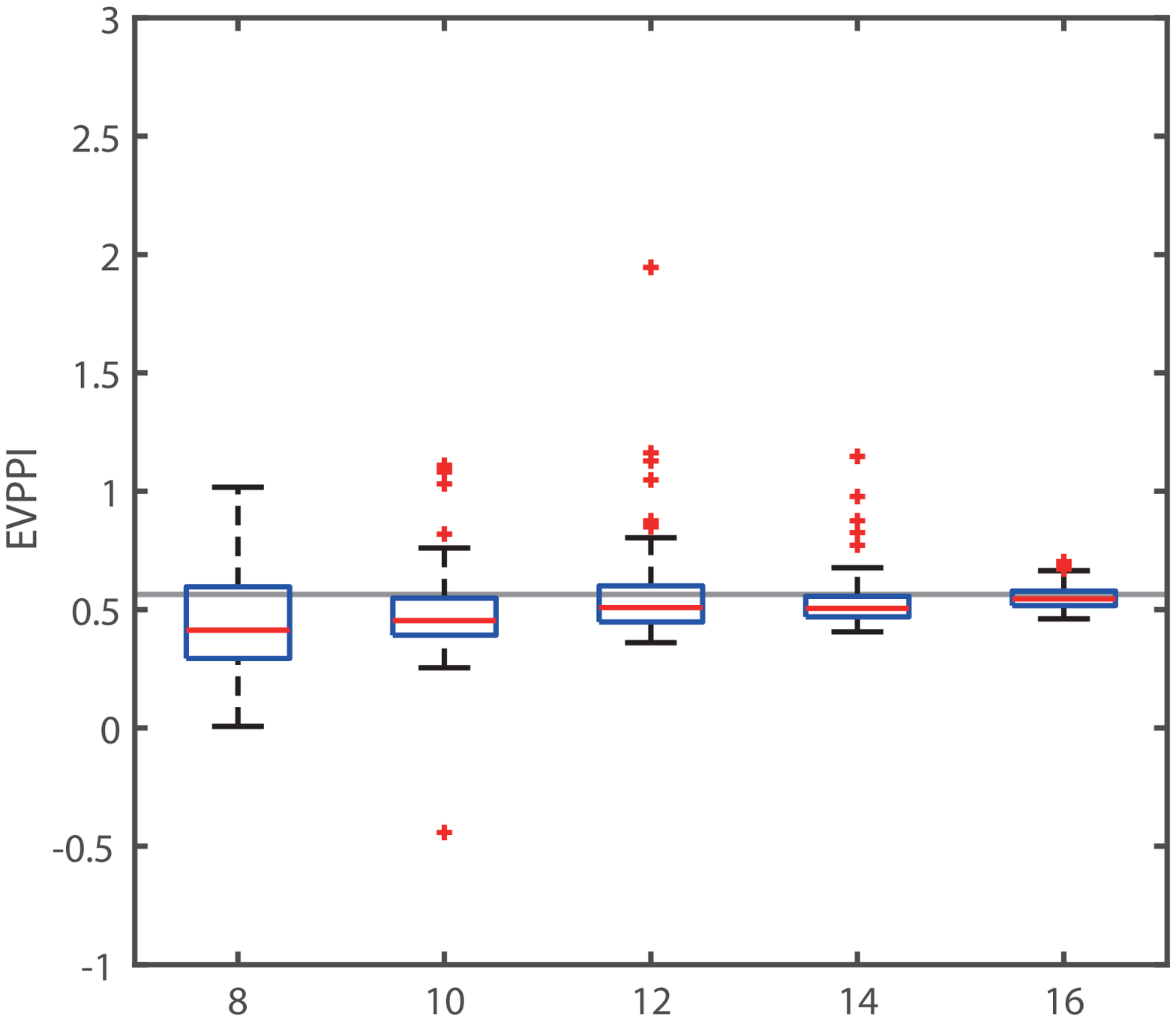}\\
\includegraphics[width=0.32\textwidth]{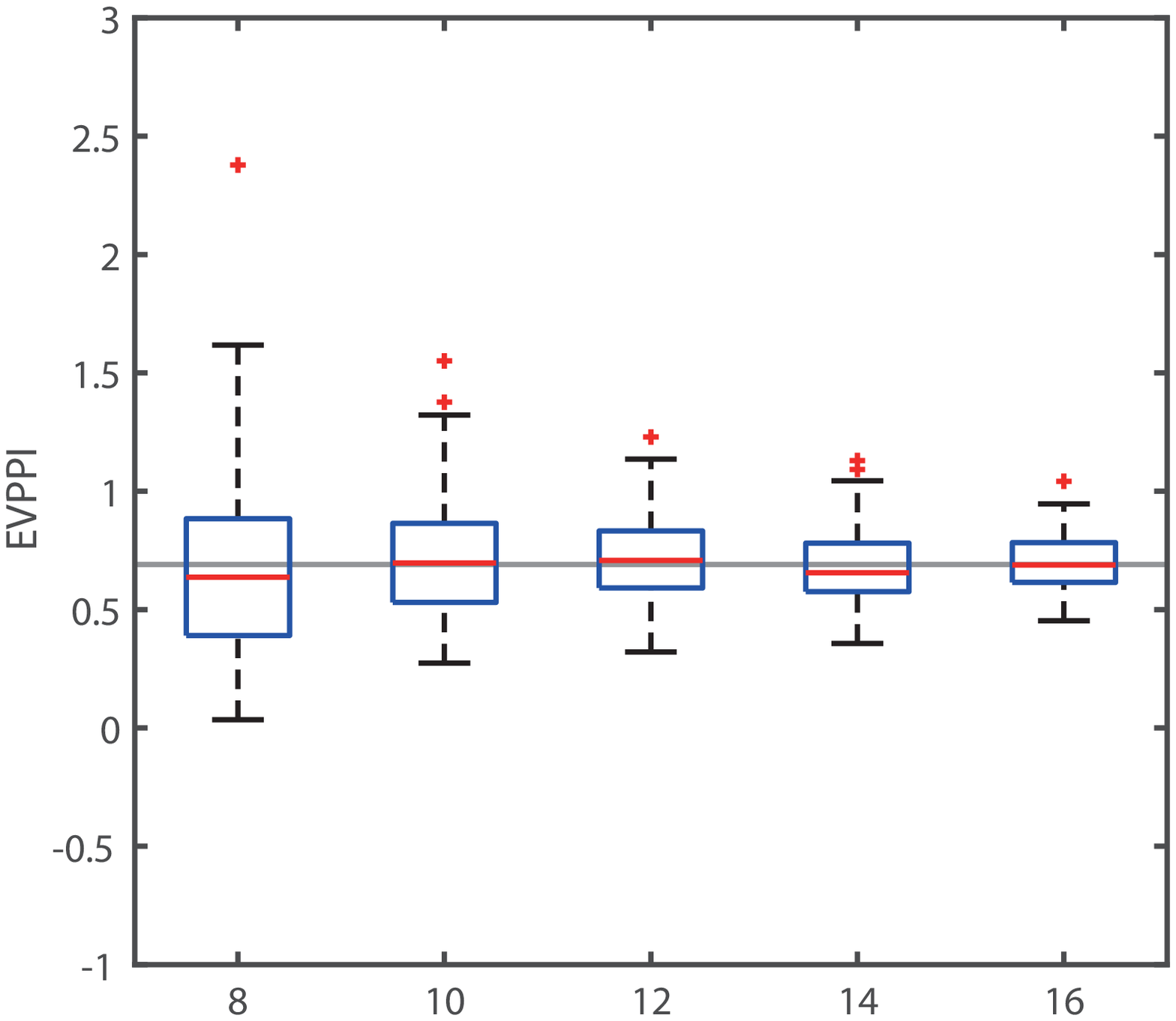}
\includegraphics[width=0.32\textwidth]{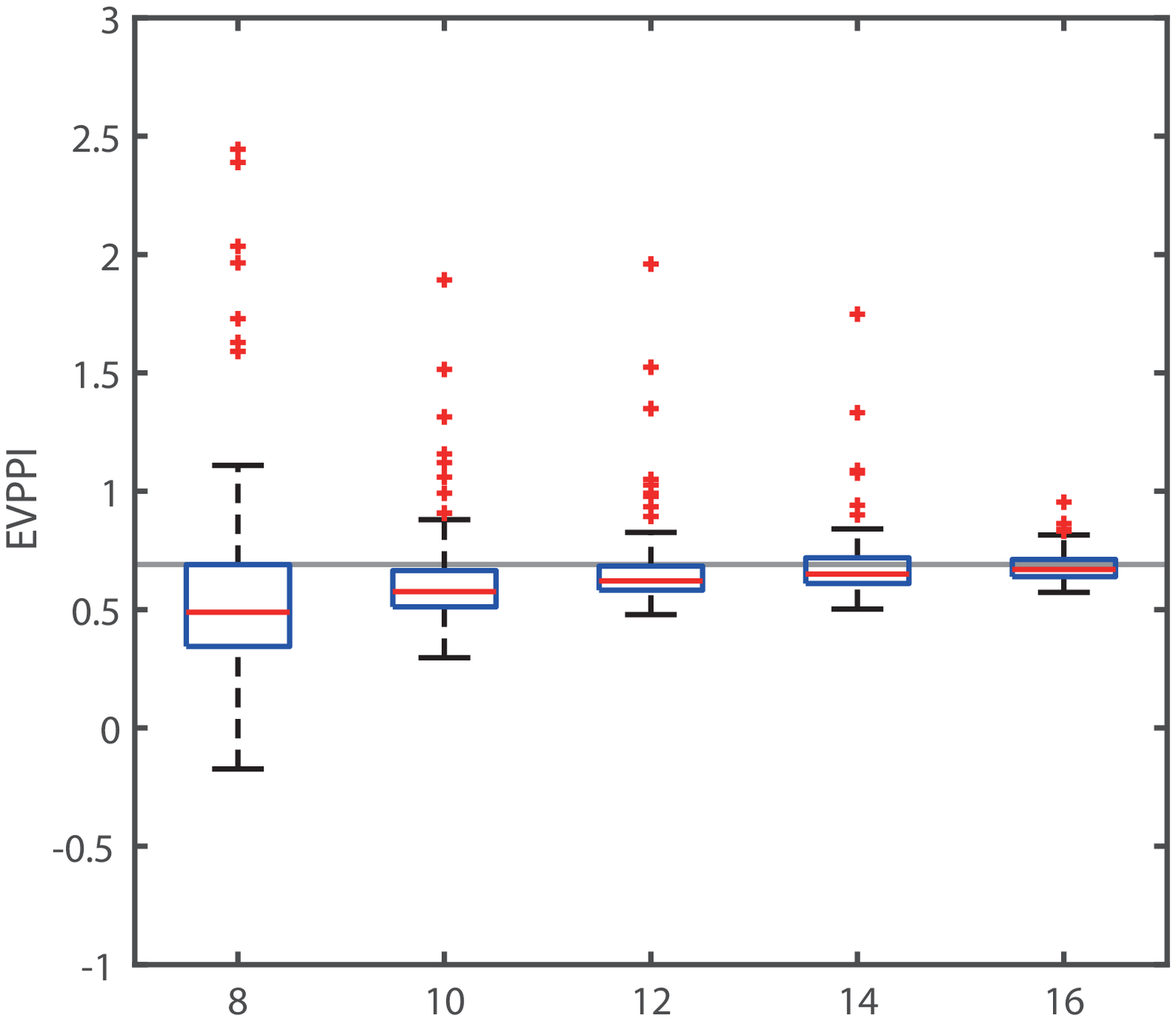}
\includegraphics[width=0.32\textwidth]{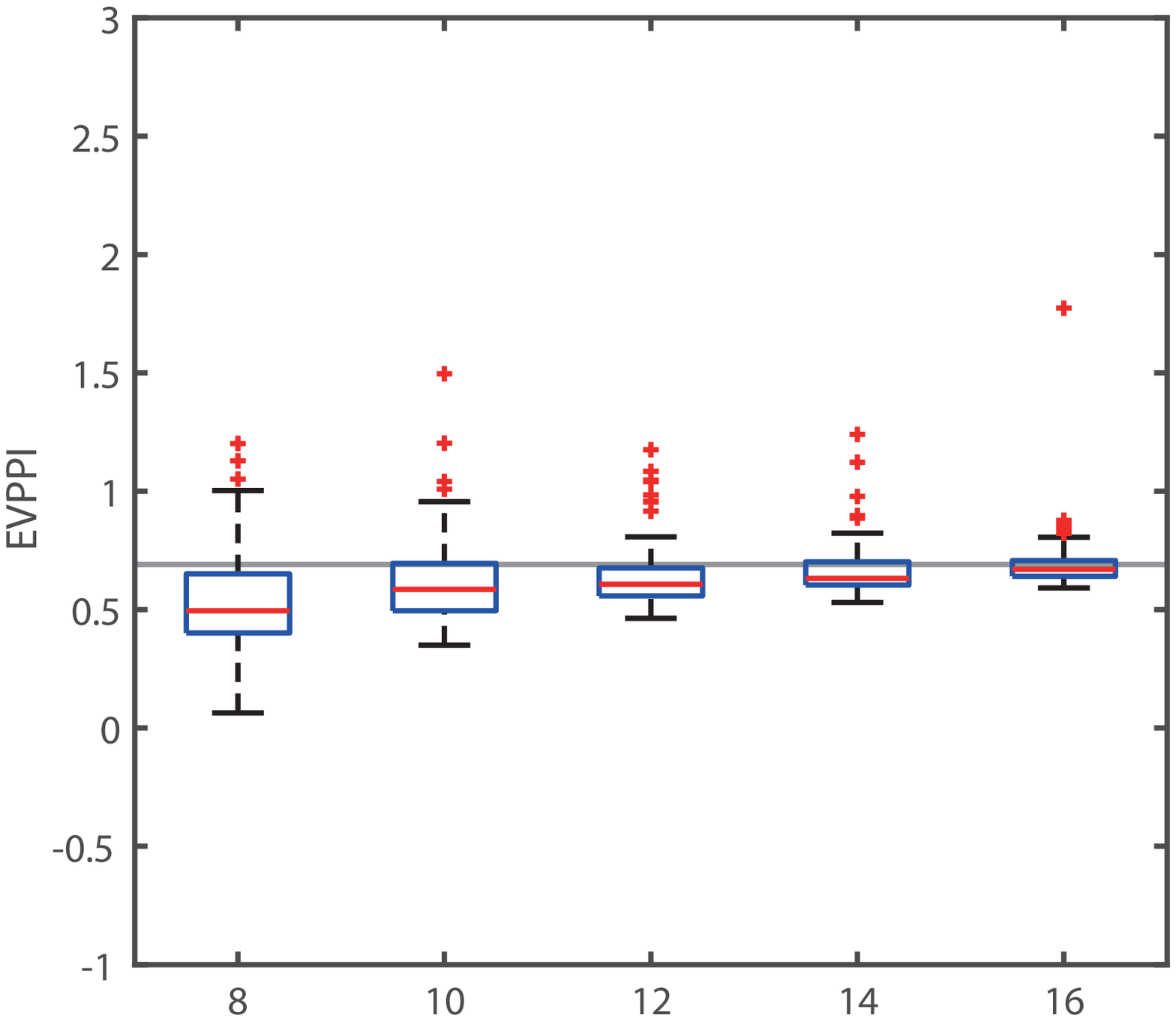}\\
\includegraphics[width=0.32\textwidth]{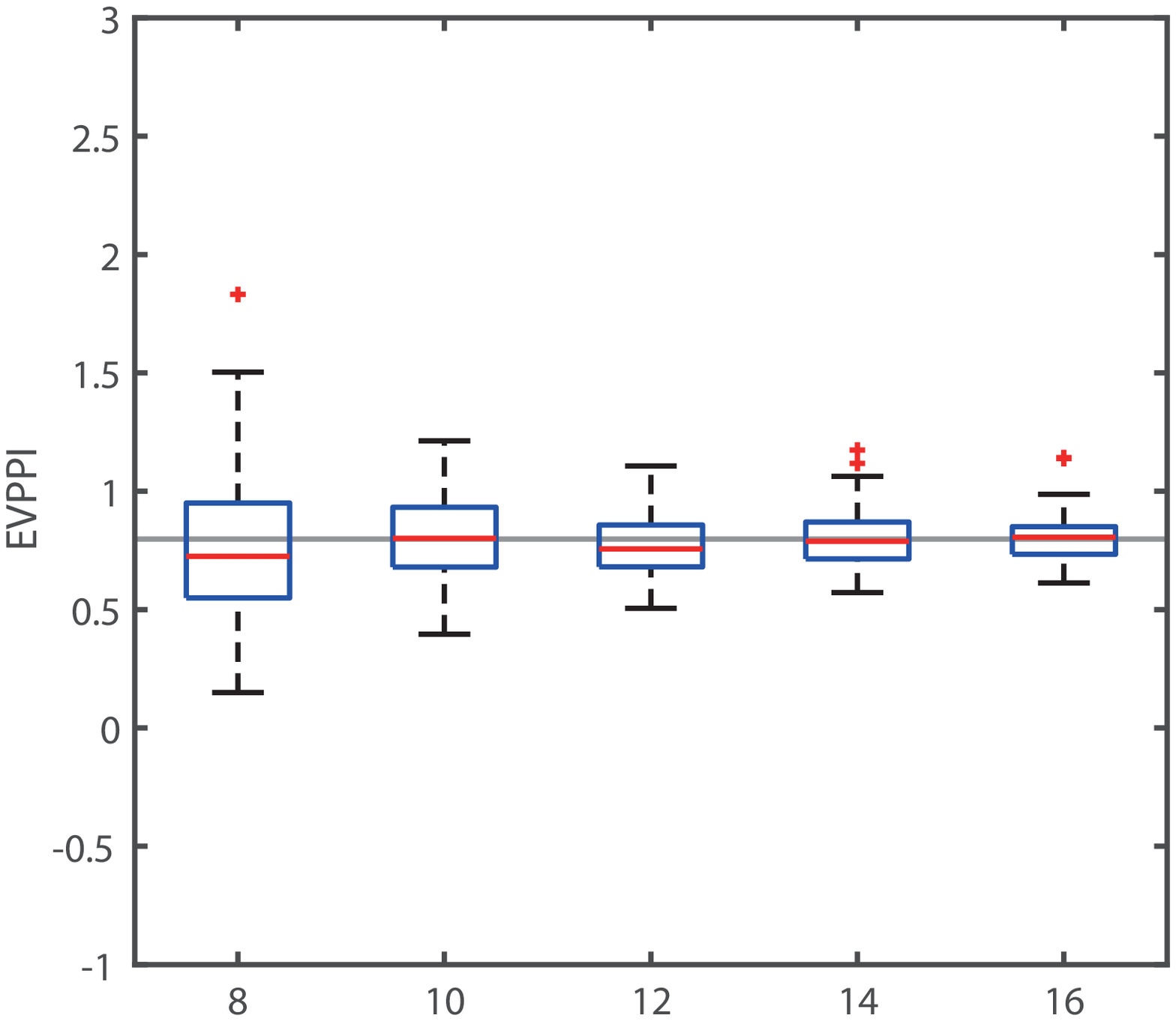}
\includegraphics[width=0.32\textwidth]{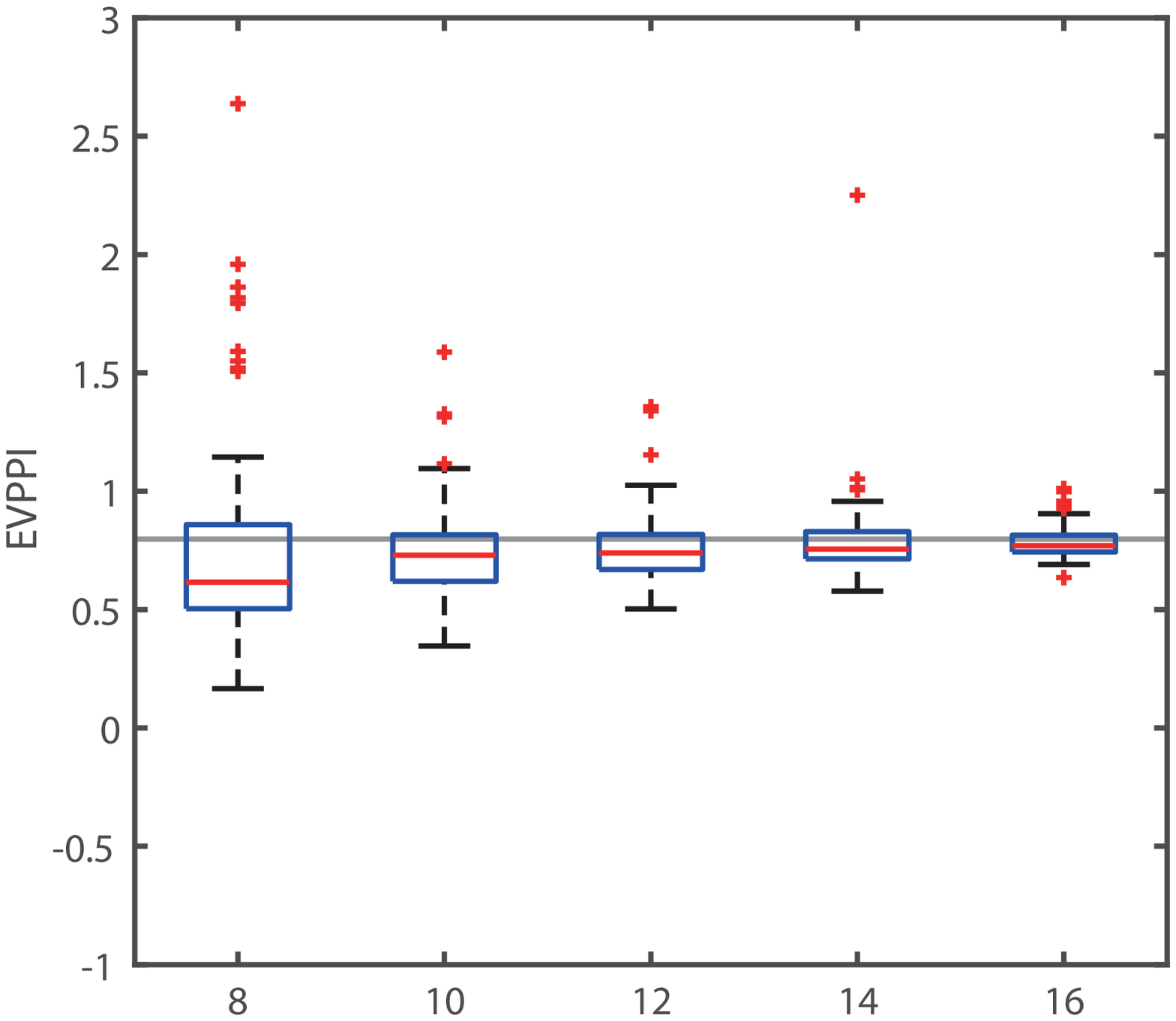}
\includegraphics[width=0.32\textwidth]{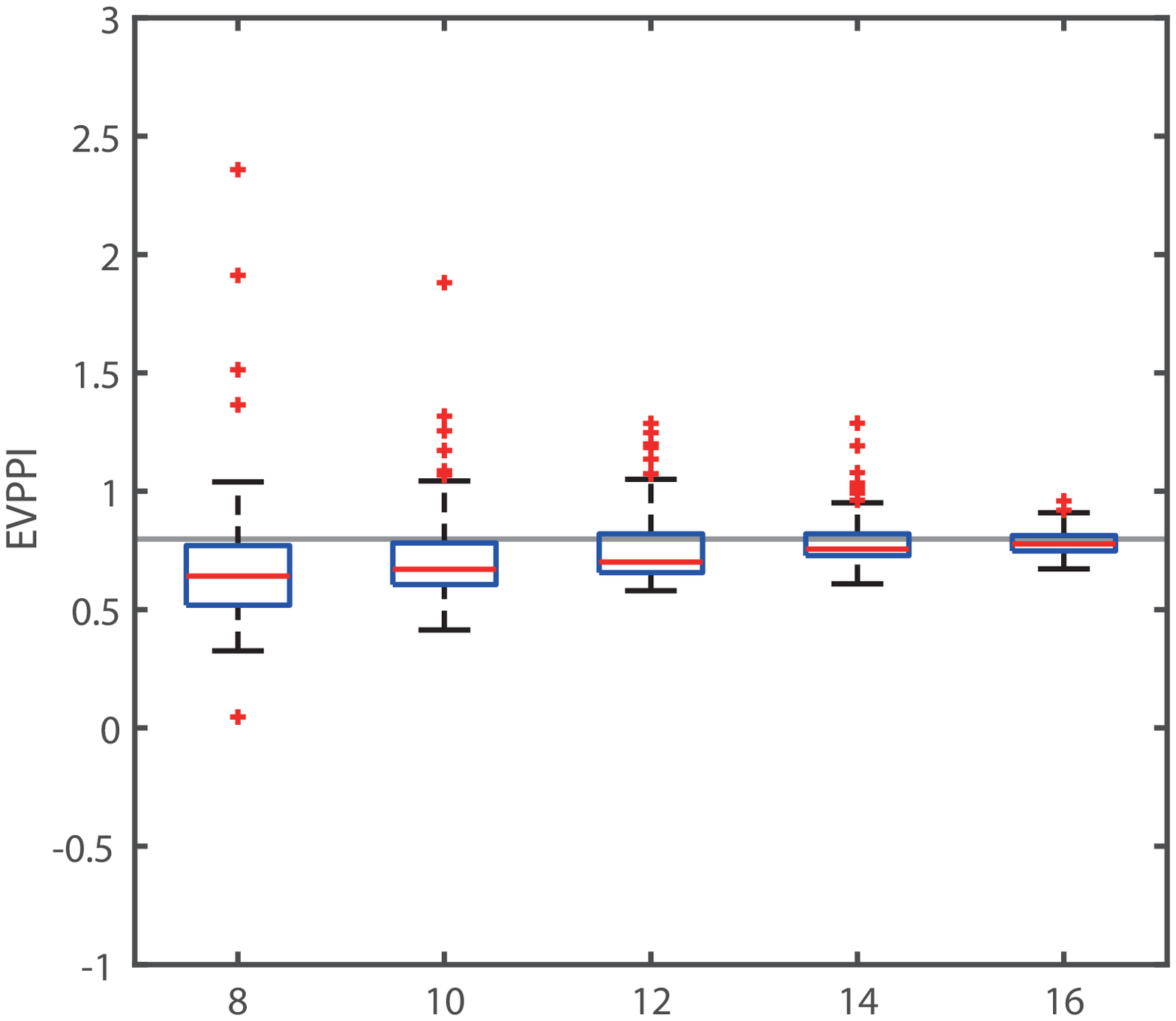}
\caption{Boxplots for 100 independent EVPPI computations on $X^{(1)}=X_1,(X_1,X_2), (X_1,X_2,X_3),(X_1,X_2,X_3,X_4)$ (from upper to lower) by the naive nested Monte Carlo estimator (left), the single term estimator (middle), and the coupled sum estimator (right) with the total computational budgets $C=2^8,\ldots,2^{16}$.}
\label{fig:evppi}
\end{center}
\end{figure}

Figure~\ref{fig:evppi} compares the boxplots of the EVPPI computations on $X^{(1)}=X_1,(X_1,X_2), (X_1,X_2,X_3),(X_1,X_2,X_3,X_4)$ (from upper panels to lower panels) obtained by three estimators as functions of $C=2^m$ with $m=8,10,\ldots,16$.
It is obvious that the convergence behavior of the nested Monte Carlo estimator is much worse than that of the Monte Carlo estimator used for computing the EVPI, as can be expected from the heuristic argument in Subsection~\ref{subsec:nested_MC}.
On the other hand, the convergence behaviors of our estimators do not differ so much whether they are used for computing either the EVPI or the EVPPI, and the length of the boxes for both of our estimators decays to 0 much faster than that for the nested Monte Carlo estimator.
Hence, as can be seen, both of our estimators give more accurate results than that of the nested Monte Carlo estimator as $C$ increases.
In practice, we recommend to use the coupled sum estimator since there are several outliers with large EVPPI values found in case of the single term estimator.

\begin{thebibliography}{99}
\bibitem{AG16}  Andrad\'ottir, S., Glynn, P.~W,: Computing Bayesian means using simulation, ACM Trans. Model. Comput. Simul. {\bf 26}, 2, Article No.~10 (2016).
\bibitem{BSLL14}  Bates, M.~E., Sparrevik, M., Lichy, N., Linkov, I.: The value of information for managing contaminated sediments, Environ. Sci. Technol. {\bf 48}, 9478--9485 (2014).
\bibitem{BGMPW08}  Bickel, J.~E., Gibson, R.~L., McVay, D.~A., Pickering, S., Waggoner, J.: Quantifying the reliability and value of 3D land seismic, SPE Res. Eval. \& Eng. {\bf 11}, 832--841 (2008).
\bibitem{BG15}  Blanchet, J.~H., Glynn, P.~W.: Unbiased Monte Carlo for optimization and functions of expectations via multi-level randomization, In: Proc. 2015 Winter Simulation Conference (2015).
\bibitem{BBL09}  Bratvold, R.~B., Bickel, J.~E., Lohne, H.~P.: Value of information in the oil and gas industry: past, present, and future, SPE Res. Eval. \& Eng. {\bf 12}, 630--638 (2009).
\bibitem{BKOC07}  Brennan, A., Kharroubi, S., O'Hagan, A., Chilcott, J.: Calculating partial expected value of perfect information via Monte Carlo sampling algorithms, Med. Decis. Making {\bf 27}, 448--470 (2007).
\bibitem{BHR15}  Bujok, K., Hambly, B.~M., Reisinger, C.: Multilevel simulation of functionals of Bernoulli random variables with application to basket credit derivatives, Methodol. Comput. Appl. Probab. {\bf 17}, 579--604 (2015).
\bibitem{Claxton99}  Claxton, K.: Bayesian approaches to the value of information: implications for the regulation of new health care technologies, Health Econ. {\bf 8}, 269--274 (1999).
\bibitem{CO08}  Coyle, D., Oakley, J.~E.: Estimating the expected value of partial perfect information: a review of methods, Eur. J. Health Econ. {\bf 9}, 251--259 (2008).
\bibitem{DTSB96}  Dakin, M.~E., Toll, J.~E., Small, M.~J., Brand, K.~P.: Risk-based environmental remediation: Bayesian Monte Carlo analysis and the expected value of sample information, Risk Analysis {\bf 16}, 67--79 (1996).
\bibitem{Delquie08}  Delqui\'{e}, P.: The value of information and intensity of preference, Decision Analysis {\bf 5}, 129--139 (2008).
\bibitem{FH98}  Felli, J.~C., Hazen, G.~B.: Sensitivity analysis and the expected value of perfect information, Med. Decis. Making {\bf 18}, 95--109 (1998).
\bibitem{Giles08}  Giles, M.~B.: Multilevel Monte Carlo path simulation, Operations Research {\bf 56}, 607--617 (2008).
\bibitem{Giles15}  Giles, M.~B.: Multilevel Monte Carlo methods, Acta Numer. {\bf 24}, 259--328 (2015).
\bibitem{Heinrich98}  Heinrich, S.: Monte Carlo complexity of global solution of integral equations, J. Complexity {\bf 14}, 151--175 (1998).
\bibitem{Howard66}  Howard, R.~A.: Information value theory, IEEE Trans. Syst. Sci. Cybern. {\bf 2}, 22--26 (1966).
\bibitem{MAPMJRW14}  Madan, J., Ades, A.~E., Price, M., Maitland, K., Jemutai, J., Revill, P., Welton, N.~J.: Strategies for efficient computation of the expected value of partial perfect information, Med. Decis. Making {\bf 34}, 327--342 (2014).
\bibitem{MMW99}  Mak, W.-K., Morton, D.~P., Wood, R.~K.: Monte Carlo bounding techniques for determining solution quality in stochastic programs, Operations Research Letters {\bf 24}, 47--56 (1999).
\bibitem{NGTS16}  Nakayasu, M., Goda, T., Tanaka, K., Sato, K.: Evaluating the value of single-point data in heterogeneous reservoirs with the expectation-maximization algorithm, SPE Econ. \& Mgmt. {\bf 8}, 1--10 (2016).
\bibitem{Oakley09}  Oakley, J.~E.: Decision-theoretic sensitivity analysis for complex computer models, Technometrics {\bf 51}, 121--129 (2009).
\bibitem{OBTC10}  Oakley, J.~E., Brennan, A., Tappenden, P., Chilcott, J.: Simulation sample sizes for Monte Carlo partial EVPI calculations, J. Health Econ. {\bf 29}, 468--477 (2010).
\bibitem{Raiffa68}  Raiffa, H.: Decision Analysis: Introductory Lectures on Choices under Uncertainty. Addison-Wesley Publishing Company, Massachusetts (1968).
\bibitem{RG12}  Rhee, C.-H., Glynn, P.~W.: A new approach to unbiased estimation for SDEs, In: Proc. 2012 Winter Simulation Conference (2012).
\bibitem{RG15}  Rhee, C.-H., Glynn, P.~W.: Unbiased estimation with square root convergence for SDE models, Operations Research {\bf 63}, 1026--1043 (2015).
\bibitem{SBZNM13}  Sadatsafavi, M., Bansback, N., Zafari, Z., Najafzadeh, M., Marra, C.: Need for speed: an efficient algorithm for calculation of single-parameter expected value of partial perfect information, Value Health {\bf 16}, 438--448 (2013).
\bibitem{SWR89}  Samson, D., Wirth, A., Rickard, J.: The value of information from multiple sources of uncertainty in decision analysis, Eur. J. Oper. Res. {\bf 39}, 254--260 (1989).
\bibitem{Sato11}  Sato, K.: Value of information analysis for adequate monitoring of carbon dioxide storage in geological reservoirs under uncertainty, Int. J. Greenh. Gas Control {\bf 5}, 1294--1302 (2011).
\bibitem{SO13}  Strong, M., Oakley, J.~E.: An efficient method for computing single-parameter partial expected value of perfect information, Med. Decis. Making {\bf 33}, 755--766 (2013).
\end{thebibliography}
\end{document}